\documentclass[12pt,preprint]{aastex}
\newcommand{\Moperyr}{M{\rm $_{\odot}~yr^{-1}$}}

\newcommand{\msol}{M{$_{\odot}$}}
\newcommand{\Msol}{M{$_{\odot}$}}
\newcommand{\lsol}{L{$_{\odot}$}}

\newcommand{\kms}{km~s{$^{-1}$}}
\newcommand{\hii}{H{\sc ii}}
\newcommand{\Hii}{H{\sc ii}}

\newcommand{\Feii}{Fe~{\sc ii}}

\def\c18o{C$^{18}$O($1\rightarrow 0$)}

\begin{document}

\slugcomment{Accepted by the Astronomical Journal, Jan 2005}

\shorttitle{Merging Massive Stars}
\shortauthors{Bally}

\title{ The Birth of High Mass Stars:  \\
        Accretion and/or Mergers?}

\author{ John Bally}
\smallskip
\affil{ Center for Astrophysics and Space Astronomy, and \\
        Department of Astrophysical and Planetary Sciences  \\
        University of Colorado, Campus Box 389, Boulder, CO 80309-0389}
\email{bally@casa.colorado.edu}

\author{and}

\author{Hans Zinnecker}
\smallskip
\affil{ Astrophyskalisches Institut Potsdam, An der Sternwarte 16, \\
14482 Potsdam, Germany}
\email{hzinnecker@aip.de}

\begin{abstract}
The observational consequences of the merger scenario for massive star
formation are explored and contrasted with the gradual accumulation of
mass by accretion.  In high density proto-star clusters,  envelopes and
disks provide a viscous medium which can dissipate the kinetic energy
of passing stars, greatly enhancing the probability of capture.
Protostellar mergers may produce high luminosity infrared flares lasting
years to centuries followed by a luminosity decline on the Kelvin-Helmholtz
time-scale of the merger product. Mergers may be surrounded by thick tori of 
expanding debris,  impulsive wide-angle outflows,  and shock induced maser 
and radio continuum emission. Collision products are expected to have 
fast stellar rotation and a large multiplicity fraction.  Close encounters 
or mergers will produce circumstellar debris disks with an orientation 
that differs form that of a pre-existing disk.  Thus, massive stars growing 
by a series of mergers may produce eruptive outflows with random 
orientations; the walls of the resulting outflow cavities may be observable 
as filaments of dense gas and dust pointing away from the massive star.  
The extremely rare merger of two stars close to the upper-mass end of 
the IMF may be a possible pathway to hypernova generated gamma-ray bursters.  
In contrast with the violence of merging, the gradual growth of massive 
stars by accretion is likely to produce less infrared variability,  
relatively thin circumstellar accretion disks which maintain their orientation, 
and collimated bipolar outflows which are scaled-up versions of those 
produced by low-mass young stellar objects.  While such accretional growth 
can lead to the formation of massive stars in isolation or in loose clusters, 
mergers can only occur in high-density cluster environments.   It is proposed 
that the outflow emerging from the OMC1 core in the Orion molecular cloud 
was produced by a protostellar merger that released between 
$10^{48}$ to $10^{49}$ ergs less than a thousand years ago.
\end{abstract}


\section{Introduction}

The birth of massive stars remains one of the outstanding problems
in star formation (Stahler, Palla, \& Ho, 2000; Larson 2003).
Stars form from the gravitational collapse of turbulent molecular 
cloud cores (Elmegreen \& Scalo 2004; Mac Low \& Klessen 2004).
There are two competing theories of massive star birth.  The traditional 
view is that massive stars form in a quasi scaled-up version of 
low-mass star formation inside dense, turbulent,  high-pressure 
hot-cores (Beech \& Mitalas 1994; Bernasconi \& Maeder 1996;
Behrend \& Maeder 2001; McKee \& Tan 2002; Tan \& McKee 2002; Yorke 2004).
Alternatively, it has been suggested that massive stars form from
the merging of lower mass protostars in high-density proto-star clusters
(Bonnell, Bate, \& Zinnecker 1998; Stahler et al. 2000; Bonnell 2002;
Zinnecker \& Bate 2002).     In this paper, we explore observational 
approaches which may distinguish between the two formation scenarios.

There are several problems with the scaled-up version of the
standard star formation scenario.  Following the pioneering studies
of Kahn (1974) and Yorke \& Kr\"ugel (1977), Wolfire \& Cassinelli 
(1986, 1987) argued that for a normal interstellar gas and dust mixture 
and stars more massive than about 10 to 20 \msol ,  radiation pressure
may stop accretional growth.  However, as the intense radiation
field dissociates molecules, evaporates ice mantles,  and sputters
grains, the opacity of the accreting material can be greatly reduced.
Nevertheless, radiation pressure tends to make accretion flows highly
unstable and for the most luminous stars, even dust free envelopes
can be blown away.   Additionally, Lyman continuum radiation
contributes to the reversal of inflow into outflow (Larson \& Starrfield 1971).   
Some of these difficulties can by circumvented by accretion from an envelope
onto a circumstellar disk followed by accretion from the disk onto
the star (Nakano et al. 1995; Jijina \& Adams 1996, Yorke \& 
Sonnhalter 2002; Yorke 2002).  Furthermore, as shown by McKee \& 
Tan (2002), accretion flows can overwhelm radiation pressure in
high-density and high-pressure cloud cores.  Nevertheless, observations
indicate that many massive stars are born in clusters in which nearby
objects can strongly alter the physical conditions from those expected
for isolated star formation.   Mutual interactions in such dense
proto-clusters may be inevitable and merging may be an important
process in some very dense star forming environments.

Predictions of the physical consequences of both accretion and merging
are needed to establish a set of observational criteria that
can discriminate between the models.  While the behavior of
stars forming by accretion are reasonably well understood, the
consequences of merging have not been fully investigated.  Therefore, 
in this paper, we explore possible consequences of protostellar merging
sibling stars (Bonnell, Bate, Zinnecker 1998).   In the next section 
(Section 2), the relevant physics of massive star formation is
reviewed.  In Section 3, plausible observable consequences are 
explored and we speculate that gamma-ray bursts may originate as a 
consequence of the merging of two massive stars in a tight binary system.
In Section 4, the properties of the nearest massive star 
forming region, the OMC1 cloud core located behind the Orion nebula, 
are interpreted as the result of a recent merger.

\section{High-Mass Star Formation Theories:  Accretion vs. Mergers}

In this section we review the salient features of the two
competing high-mass star formation models relevant to observations
designed to distinguish between them.

\subsection{The Accretion Model}

In the standard model of star formation, massive stars form in
a manner similar to low-mass stars but with very high accretion rates
(see reviews in Protostars \& Planets IV, eds. Mannings, Boss, and Russell 2000).
To assemble a $\sim$100 \Msol\ star in $10^{5}$ yr requires an 
average accretion rate of $\dot M = 10^{-3}$ \Moperyr .
For an isothermal sphere with an $r^{-2}$ radial density profile, the
standard model of inside-out collapse predicts a mass accretion rate
$\dot M \sim c_s^3 / G$ where $c_s$ is the effective sound speed, and
$G$ is Newton's gravitational constant (e.g. Hartmann 1998).  An
accretion rate $\dot M = 10^{-3}$ \Moperyr\ requires $c_s$ = 1.9 \kms\
which corresponds to a temperature of about 1,200 K for molecular
gas (assuming a mean molecular weight of $\mu = 2.7$
since hydrogen is expected to be molecular).
If a cloud core contains 100 \msol\ and collapses to form a
massive star in $\tau = 10^5$ years, then the initial radius of
the core must be about $R = 5 \times  10^{17}$ cm,
its average $H_2$ density $n(H_2) = 3 \times 10^4$ cm$^{-3}$, the
average pressure  $P / k = 3 \times 10^8$ cm$^{-3}$~K, and its
average column density  $N(H_2) = R \times n(H_2) = 3 \times 10^{23}$
$\rm cm^{-2}$ corresponding to about $A_V \sim 200$ magnitudes
(Tan \& McKee 2002; McKee \& Tan 2002).
Thus, growth by accretion requires very high pressure cores, large
effective sound speeds, and high column densities.

It is possible that relatively isolated massive stars
can be produced by radiation driven implosion
(Klein, Sandford, \& Whitaker 1983; Ho, Klein, \& Haschick 1986).
Several ultra-compact HII regions such as M17 UC1 are embedded in
dense cores adjacent to ionization fronts that have partially enveloped
them (e.g. Nielbock et al. 2001; Chini et al. 2000;
Hanson, Howarth, \& Conti 1997).  The increased pressure generated
by adjacent HII regions can compress the cloud core and trigger
the large accretion rates needed to form a high-mass object
(Bertoldi 1989; Kessel-Deynet \& Burkert 2003).  Such a scenario may
be responsible for the large accretion disk recently found surrounding
a massive protostar at the periphery of the \hii\ region M17
(e.g. Chini et al. 2004). 

A major difference between high- and low-mass star formation is
that the pre-main sequence contraction time-scale is shorter than the
accretion time-scale for $>$ 10 \Msol .  Thus, 
while low-mass stars always emerge from their natal cocoons as 
pre-main sequence objects, massive stars reach the zero-age main 
sequence and maturity while still embedded within their parent clouds, 
and while still accreting.

The effects of the forming star's radiation pressure, abundant soft-UV,
and ionizing Lyman continuum radiation must be included in models of
its birth.  For a normal interstellar gas and dust mixture,
radiation pressure can reverse inflow and prevent growth for massive stars.
However, as the intense radiation field dissociates molecules, evaporates ice
mantles,  and sputters grains, the opacity of the accreting
material may be reduced greatly.   Even with grain destruction,
radiation pressure may halt the growth of massive stars by accretion
for masses greater than 20 \msol\ (Wolfire \& Cassinelli 1986, 1987).
Tan \& McKee (2002) and McKee \& Tan (2002) argue that the high
pressure of a collapsing massive cloud core can overwhelm the radiation
pressure of a massive star forming at its center.  For a star with
a luminosity $L_* = 10^5$ \lsol , the radiation pressure is
$P_{rad} / k = L_* / 4 \pi c k R^2 = 3 \times 10^7$ cm$^{-3}$~K,
where R is the stellar radius, c is the speed of light, and
k is Boltzman's constant.  $P_{rad}$ is about an order of magnitude 
lower than the average pressure in the core discussed above.   
Therefore, the growth of an \Hii\ region can also stop accretion.
But, Keto (2002) has shown that the gravity of the central star
can confine an the HII region and bottle-up the destructive effects of Lyman
continuum radiation field if its size is less than about
$r_{II} = GM / c_{II}^2$ where $c_{II} \approx$ 10 \kms is the sound
speed in the \Hii\ region.  The ionized zone surrounding a massive star 
can remain hyper-compact if there is a reservoir of gas with a density 
$n(H) > (3 L(LyC) / 4 \pi \alpha_B )^{1/2} r_{II}^{-3/2}$ 
within a distance $r_{II}$ of the star.   If an accretion flow 
or a circumstellar disk can provide this reservoir, the \Hii\ region
can remain confined and the destructive effects of UV radiation 
can be mitigated.  In a a disk-confined \Hii\ region, 
some UV may escape and ionize gas at large distances orthogonal to the 
disk plane.   However, an accretion flow in the disk plane can 
remain shielded.  In rapidly rotating stars, the poles are hotter 
than the equator.  Thus,  most of the ionizing radiation is beamed 
towards the poles, reducing its destructive effects in the equatorial 
plane.   Yorke \& Sonnhalter (2002) modeled the formation of 
massive stars with the inclusion of these effects, demonstrating that 
direct accretion can produce stars up to 20 to 30 \Msol\ by accretion 
from a disk.  But, how do the most massive stars form?

Henning et al. (2000) found a correlation between outflow mass-loss rate 
from high mass sources with their source luminosity.  As
a possible interpretation of this correlation,  Behrend \& Maeder (2001)
suggest that the accretion rate onto massive stars may increase with 
time, contrary to the observed behavior of isolated low-mass stars.
This feature can be incorporated into the direct accretion picture 
of massive star birth in several ways.  
First, 
the accretion rate will tend to increase with time if the slope 
of the parent cloud core density profile is shallower than 
the $r^{-2}$ density profile of an isothermal sphere.   
Second, 
as the mass of a star increases, its Bondi accretion radius grows 
so that the star can accumulate mass at an increased rate from 
a given density environment.  This effect, known as 
``competitive accretion'', can lead to run-away growth of 
the most massive objects in a forming cluster (Zinnecker 1982;
Bonnell et al. 2001a, b).  
Third, 
massive objects tend to settle into the center of a forming cluster
where the density of gas, and hence the accretion rate,  
is higher. 
However, Beuther et al. (2002) find evidence for steady outflow rates
and argue that massive stars may indeed form via a scaled-up version 
of low mass star formation.

\subsection{Mergers}

Why consider alternative models for massive star birth? First,
observations demonstrate that massive stars tend to form in clusters
rather than in isolation (e.g. Lada \& Lada 2003; De Wit et al. 2004).
The central densities of young clusters sometimes
exceed $10^5$ stars per cubic parsec (e.g. the Trapezium cluster
in Orion --  Henney \& Arthur 1998; IRS5 in W3 -- Claussen et al. 1994).  
In such dense environments,
perturbations by and interactions with sibling stars must be considered.
Second, young OB associations and clusters expand and 5 to 30\% of their
members are ejected as high velocity stars (Gies 1987; Stone 1991; 
Kroupa 1995; Hoogerwerf, de Bruijne, \& de Zeeuw 2000).  Third, 
some outflows from very massive stars are poorly collimated 
(e.g. OMC1 in Orion; McCaughrean \& Mac Low 1997; Kaifu et al. 2000) 
and look like the result of an explosion (Allen \& Burton 1993).  Other 
outflows from massive proto-stars show evidence of abrupt flow-axis 
reorientation (e.g. IRAS 20126+4104 - Shepherd et al. 2000).  Fourth, 
massive stars have a larger binary companion frequency (Mermilliod \& 
Garcia 2001; Zinnecker \& Bate 2002)  and faster rotation 
rates than low-mass stars.  As discussed below, these trends
can be interpreted as evidence that interactions with sibling stars
may play important roles in the formation of some massive stars.
Models of galactic collisions (e.g Toomre \& Toomre 1972) and interactions
between stars and protostars (e.g. Soker et al. 1987; Benz \& Hills 1992;
Clarke \& Pringle 1993; Hall, Clarke, \& Pringle 1996; Larwood 1997;
Boffin et al. 1998; Watkins et a. 1998a, b; Portegies Zwart et al. 1999;
2002; Lombardi et al. 2003) provide insights into the relevant 
physics of such interactions.

Bonnell, Bate,  \& Zinnecker (1998) proposed that massive
stars form by a combination of competitive accretion and merging
in forming star clusters.  For star--star collisions to occur
at a significant rate, a cluster requires a number density of about
$10^8$ stars pc$^{-3}$ for low-mass stars.  However, for 10 \msol
stars with a velocity dispersion of 5 to 10 \kms , merging may
occur at densities as low as $10^6$ stars pc$^{-3}$.
Though such high densities are not found
in any observed star cluster, Bonnell et al. (1998) and Stahler
et al. (2000) argue such extreme conditions may exist for a brief
period of time during the highly embedded and obscured gravitational
collapse and fragmentation phase of cluster forming cloud cores.
As discussed next, gravitational focusing and dissipation by
circumstellar matter can greatly enhance the merger rate in
forming clusters,  lowering the required density of stars during such a phase
(e.g. Bally 2002; Bonnell 2002, Zinnecker \& Bate 2002).

\subsubsection{Gravitational Focusing}

The cross-section for interactions among stars can be orders
of magnitude larger than the projected area of a star because the
typical velocity dispersion in young clusters is much smaller
than the escape speed from a typical stellar surface.  Consider a
star of mass $m$ approaching a star with mass
$M$ at a relative velocity $v$ (which at large separation is roughly
given by the cluster velocity dispersion).   The relative velocity
at periastron is given by
$$
v_{p}^2 = v^2 + {{ 2 G (m + M) } \over {r_p}}
$$
where $r_p$ is the periastron separation.
From the conservation of specific angular momentum
it follows that $ b v_{clus} = r_{p} v_{p}$ where $b$ is the
impact parameter.  When the separation between the centers-of-mass
of the stars at perihelion is less than the sum of the stellar radii,
the interaction results in a collision that may lead to the formation
of a binary or possibly to a merger.  The cross-section for
a grazing collision is given by
$$
 \sigma_{f} = \pi r_f^2 = \pi (r_m + r_M)^2 \Bigl[ 1 + {{2 G (m + M)} \over
    { (r_m + r_M) v_{0}^2  }} \Bigr] \approx 
  \pi (r_m + r_M)^2 \Bigl[ 1 + {{ v_{esc}^2 } \over {4 v_{clus}^2}} \Bigr] 
$$
where the subscript $f$ refers to gravitational focusing,
$r_m$ and $r_M$ are the radii of the stars,
$v_{esc} = 2 G(m + M)/(r_m + r_M)$ 
is the gravitational escape speed from the combined mass at periastron 
separation, and $v_{0} \approx 2 v_{clus}$ is the 
relative speed of the two stars.  Thus, gravitational focusing 
increases the collision cross-section by roughly a factor
$f_G = (1 + v_{esc}^2 / 4 v_{clus}^2) \approx (v_{esc} / 2 v_{clus})^2$.
For the Orion Nebula cluster, $v_{clus} \approx 1.5$ \kms\ 
(van Altena et al. 1988).  Using $v_{esc} \approx 300$ \kms\ 
corresponding to a slightly bloated pre-main-sequence star, 
gravitational focusing enhances interaction cross-sections 
by about a factor of $10^4$.

\subsubsection{Disk-Assisted Protostellar Capture}

Massive circumstellar disks and/or envelopes can increase the
cross-section for stellar capture by additional orders of magnitude.  
Disk mediated interactions were considered as a possible binary formation 
mechanism by Clarke \& Pringle (1991; 1993), and Heller 
(1995).  These authors considered collision partners having similar 
masses and concluded that such interactions can at most account for 
a small fraction of the observed binary star population in young clusters.  
The youngest clusters are much denser than those considered by Clarke 
and Pringle and often contain large amounts of interstellar and 
circumstellar gas.   A careful re-assessment of the role of encounters 
during the embedded phases of young clusters is warranted.

Numerical modeling of close encounters indicates that disks
are disrupted and truncated to a new outer radius equal to
roughly $1/2$ or $1/3$ of the periastron separation 
(Clarke \& Pringle 1993; Hall, Clarke, \& Pringle 1996; 
Heller 1995; Larwood 1997; Hall et al. 1998; 
Watkins et al. 1998a,b; Boffin et al. 1998).  The ejected portion of 
the disk can absorb the incoming star's kinetic energy and facilitate 
the formation of a binary.  A star of mass $m$ can be captured by 
another star with mass $M$ that is orbited by a disk with mass $M_d$ 
if the two stars approach each other within roughly a disk radius and 
the disk has enough mass to lower the impactor's velocity to below 
escape speed (Clarke \& Pringle 1991).  This condition is satisfied 
if the (negative) binding energy of the disk is
larger than the excess kinetic energy (i.e. at infinite separation)
of the impactor star.  That is,
$$
{1 \over 2} m v_m^2  < { {G M_d M} \over { r_d }}
$$
where $v_m$ is the velocity (at large separation) of the star with mass
$m$ with respect to star $M$ ($v_m \approx v_{clus}$, the cluster velocity
dispersion) and $r_d$ is the effective half-mass radius of the disk
(the radius at which a point mass with mass $M_d$ has the
same gravitational potential energy as the disk).  Thus, circumstellar
matter-assisted capture is possible if the encounter velocity
(at large separation) is
$$
v_m < \Bigl({{2 G M} \over { r_d}} \Bigr)^{1/2} \Bigl( {{M_d} \over {m}} \Bigr)^{1/2}
= v_{esc}(r_d) \Bigl( {{M_d} \over {m}} \Bigr) ^{1/2}
$$
and periastron passage brings the impactor to within the half-mass radius
of the disk.  Thus a $M$ = 10 \msol\ star surrounded by a
$M_d$ = 0.1 \msol\ disk (1\% of the mass of its central star) having a radius
$r_d$ = 10 AU can capture a star with a mass m = 1 \msol\ if the relative
velocity is less than about 6 \kms\ and the stars pass within about a disk
radius.  Because for typical parameters, $r_f \sim$ 1 AU,
direct collisions are much less likely than disk-mediated interactions
which can occur for impact parameters as large as 100 AU.

Numerical models (e.g. Pfalzner 2003) demonstrate that interactions excite
strong spiral density waves and tidal tails which can efficiently transfer 
angular momentum and mass through the disk.  Some of the disk mass is ejected 
from the system and some is accreted onto the central star.   The impactor 
can also steal a portion of the impacted disk mass.   

Disk-assisted capture of an intruder usually results in the formation of an 
eccentric binary.  The initial orbit of the captured star tends to shrink
somewhat as the outer parts of the disk are ejected.  However, once the 
interaction truncates the disk(s) to an outer radius smaller than the 
separation of the stars during periastron passage, orbital decay stops
(Moeckel \& Bally 2005).  

In a dense cluster forming environment, two processes may lead to further
orbit evolution;  ongoing accretion from the surrounding envelope, and an 
interaction with another star (Bate, Bonnell, \& Bromm 2002; Bonnell, Bate,
\& Vine 2003; Moeckel \& Bally 2005).   Accretion adds mass to the 
system and therefore leads to orbital decay.  The resulting decrease in 
periastron separation can lead to further dissipative interaction with remnant 
disks and hardening of the binary.   Interaction with a second intruder star 
can re-configure the binary and sometimes may lead to a merger.   
Three-body encounters leading to mergers were observed in the simulations 
of Bonnell \& Bate (2002).  Numerical simulations are needed to identify 
the parameters which lead to mergers or captured binary companions.   
In the rest of this paper, we restrict our attention to those encounters 
which do lead to merging.

\subsubsection{Merger Types}

In most models of star formation, the gravitational collapse
of a pre-stellar cloud core leads to the formation of a magnetically
or rotationally supported disk;  outward angular momentum transport in the
disk then drives accretion onto a central protostar.  Thus, there
are three different types of entity in star formation; cores,
disks, and (proto)stars.  To form a star cluster, a portion of a giant 
molecular cloud must fragment before or during its collapse phase.  
Since the resulting cores may collapse at different times and evolve 
at different rates,  a forming cluster is expected to consist of a 
mixture of cores, disks, and protostars (e.g. Klessen 2001; Bate et al. 2003;
Schmeja, Klessen, \& Froebrich 2005).

In principle, six types of mutual interactions are possible among these 
three entities.  The top row of Figure~1 shows the interactions that involve 
cores.   Since cores are diffuse, mutual interactions between these 
objects will be relatively low-energy events.  Nevertheless, a massive star, 
with or without its own circumstellar disk, can potentially gain additional 
mass by merging with a passing core.  However,  the addition of more mass
to the envelope of a massive star does little to overcome potential problems 
with radiation pressure and ionization.   High velocity collisions with 
cores can strip a protostar of its own core or disk, bringing accretional 
growth to a halt (Price \& Podsiadlowski 1995).  These events are likely 
to have considerably less spectacular consequences than disk-disk, disk-star, 
or star-star interactions.   Star-star collisions (bottom row in Figure~1) 
are likely to be very energetic, but are the least likely in the star 
formation environment.  Interactions between young stars surrounded by 
circumstellar disks (middle row in Figure~1) are likely be the most 
interesting interactions in nascent clusters.  A massive protostar 
is most likely to have an encounter with the most common type of object 
in such a forming cluster -- a low-mass star, disk, or core.  
However, is a mass-segregated cluster, the most common type of encounter
may be a massive protostar with another massive or intermediate mass
star.

\subsection{Merger Energetics}

The energy released by a merger of a low-mass object of mass $m$ 
with a more massive object having mass $M$ is $ E = \eta GmM/R $ 
where $R$ is radius of the collision product and $\eta$ is a factor 
of order unity that takes into account details of the density 
distributions of the stars and the additional mass of circumstellar 
material.  For $M = M_{10} = 10$ \msol\ , $m = m_1 = 1$ \msol ,
and $R = 10^{12}$ cm, we find $E = 2.5 \times 10^{48} \eta$ ergs.
The range of energies can cover more than 6 orders of magnitude; while
the merger of two 0.1 \msol\ dwarf protostars produces only
$3 \times 10^{45}$ ergs, the much rarer merger of two 100 \msol\ stars
would release more than $10^{51}$ ergs.  {\it The estimated energies
of observed protostellar outflows lie within the range
produced by mergers.}

The most probable encounters have large impact parameters and therefore
have high angular momenta.  Such interactions can efficiently
convert impactor kinetic energy into rotational energy.  As discussed
below, some of this spin energy can be carried away by the ejection
of the outer disk and some can be carried away by an outflow emerging 
along the axis of the system.

\subsubsection{The Merger Luminosity}

Most of the gravitational potential energy 
is released when the stars merge and the peak energy 
release rate can reach values as high as
$ L \approx  (GM)^{3/2} m / 2 \pi R^{5/2} $
$ = 1.2 \times 10^{9} m_1 M_{10}^{3/2} R_{12}^{-5/2}$~\lsol\
where $R$ is the orbital radius just before merging (comparable to
the radius of the collision product) and
$R_{12}$ is in units of $10^{12}$ cm. 

The released gravitational potential energy is deposited in
the merger product and surrounding debris.  The luminosity 
of the merger product will be limited by the radiative time-scale
which can be estimated from 
$$
L_{max} \approx G \eta mM / \tau_{c} R
= 2.1 \times 10^3 m_1 M_{10} \tau_{4}^{-1} R_{12}^{-1} ~~~ (\rm L_{\odot} )
$$
(for $\eta = 1$)
where $\tau_{c}$ is the radiative cooling time of the debris cloud
surrounding the merger product and 
$\tau_{4}$ is this cooling time in units of $10^4$ years.
Though $\tau_{c}$ is likely to be a complex function of the debris geometry 
and the stellar masses involved, it can be estimated in several ways.    

The radiative cooling time of the collision product is 
likely to be comparable to its Kelvin-Helmholtz time   
given by $\tau _{K-H} \approx G M^2 / R L$.
Using the main-sequence mass-luminosity relationship, 
$L \propto M^{3.3}$, and mass-radius relationship, $R \propto M^{2/3}$
implies that the Kelvin-Helmholtz time scales as 
$\tau _{K-H} \approx 1.1 \times 10^5 M_{10}^{-2}$ years where
$M_{10}$ is the mass of the product in units of 10 \msol . 

However, immediately after its formation, the merger product is 
likely to have a photospheric radius considerably 
larger than a main-sequence star with the same mass.   
Numerical models of stellar collisions leading to the formation 
of blue stragglers in globular clusters (Lombardi et al. 2002, 
2003; Fregeau et al. 2004) indicate that merger products have 
radii up to 30 times larger than an equivalent main sequence star.
Thus, $\tau _{K-H}$ for a merger product is likely to be shorter 
than the above estimate for a main sequence star by at least an 
order of magnitude.   Thus $\tau _c$ may be in the range 
$3 \times 10^3$ to $5 \times 10^4$ years for a 10 \msol\ product
and less than 100 years for a 100 \msol\ product.

The cooling time can also be estimated from the time
required for photons to diffuse from the merger product 
and its surrounding debris field.  The diffusion time-scale 
$\tau_{diff} = (l / c) N$ where $l$ is the photon mean free
path,  $c$ is the speed of light, and $N \approx (R / l)^2$ 
is the number of scatterings the photon experiences during its 
random-walk to the surface.  Thus, 
$\tau_{diff} \approx  \kappa  M / c R$ where $\kappa$ is the
opacity per unit mass, $M$ is the mass of the merger product or
its debris field, and $R$ is the size-scale.

The merger product is likely to be surrounded by a debris 
field that has been shock heated by the merger.  This material 
will have an even shorter cooling time because it will be 
spread over a region larger than the product.  Additionally,  
debris will tend to be confined to a flattened 
structure preferentially lying in the encounter plane.
Thus, there may be two distinct radiative cooling time-scales
for a merger product.  The shorter time-scale associated with
the larger but less massive debris field, and a longer time-scale
for the merger product.  While a 10 \msol\ collision product may 
relax in $10^4$ to $10^5$ years, a 100 \msol\ remnant 
may relax in less than $10^3$ years leading to  
the peak emergent luminosities ranging from $10^4$ to nearly
$10^7$ \lsol .

Mergers are most likely to occur during the short-lived, 
high-density, embedded phase of a forming cluster.  Thus, most
of the emergent radiation is likely to be obscured and re-processed 
to far-infrared or longer wavelengths by the surrounding 
medium.

\subsubsection{Ionizing Radiation}

Stars more massive than about 15 \msol\ contract to the main-sequence
faster than they can accrete their mass in most models (e.g. Yorke 2003).
Thus, accreting or merging stars more massive than this can be treated 
as main-sequence objects with a well determined Lyman continuum luminosity.    
Their \hii\ regions can be confined by gravity when the density of the 
surrounding medium is sufficiently large and the radius of the ionized 
zone is sufficiently small to make the escape speed from this radius 
larger than the sound speed in photo-ionized plasma.   The radius, 
$r_g$, at which the escape speed from a star of mass $M$ is given by 
the sound speed, $c_{II}$, is given by $r_g = GM / c_{II}^2$.   Combining this 
with the Stromgren radius of an \hii\ region, given by 
$r_S = (3 L(LyC) / 4 \pi \alpha _B n_e^2)^{1/3}$,
where $L(LyC)$ is the Lyman continuum luminosity of the ionizing star,
$\alpha _B$ is the case-B recombination coefficient, and $n_e$ is the 
electron density, gives the critical density above which the gravitational 
field of the star can confine the plasma.  That is,  a uniform density 
\Hii\ region may be bound by the gravity of the UV-emitting star if
$n_e > (3 L(LyC)/ 4 \pi \alpha _B )^{1/2} (GM)^{-3/2} c_{II}^3 $
(e.g. Keto 2002).  Thus, a 10 \msol\ star with 
$L(LyC) = 10^{45}$ photons s$^{-1}$ has a gravitational
radius, $r_g = 90$ AU and a critical density $n_e = 6 \times 10^5$ cm$^{-3}$
above which the \hii\ region will be confined by gravity.  The 
corresponding numbers for a 100 \msol\ star with 
$L(LyC) = 10^{50}$ photons s$^{-1}$
are $r_g = 900$ AU and $n_e = 6 \times 10^6$ cm$^{-3}$.

If a core or disk with a much higher density than this critical density
penetrates to a distance $d$ of a massive star, the characteristic radius 
of the resulting \hii\ region will become comparable to $d$.  The density 
of the photo-ionized plasma can be estimated from photo-ionization equilibrium 
since to first order, recombinations in a layer of thickness comparable to 
$d$ must consume the incident flux of Lyman continuum photons.   The density 
of the photo-ionized plasma between the massive star and the dense intruding 
body will adjust itself to satisfy the Stromgren criterion, 
$n_e \approx (3 L(LyC) / 4 \pi a_B )^{1/2} d^{-3/2}$.
This plasma will be confined by gravity when $d < r_g$.  Thus, 
the collision of a massive star with an ultra-dense cloud core fragment or 
a lower-mass star surrounded by a dense disk may result in a
gravitationally confined hyper-compact \hii\ region which can smother the
star's Lyman continuum radiation field. 

\subsection{Merger Evolution}

Detailed consideration of the merger process will require
numerical modeling in which the hydrodynamics, effects of
gravity, magnetic fields, and radiation are considered carefully.
Here, we break the problem into sub-processes and speculate
about the likely behavior of the system.  If a merger does occur,
there are likely to be three distinct phases; an {\it in-spiral}
phase following the initial encounter during which the 
system evolves toward a more compact configuration, the {\it merger}
itself during which the bulk of the gravitational potential energy is
released, and the post-merger {\it outflow} phase during which some
circumstellar matter is ejected by this energy and the collision product 
settles into a new configuration.

\subsubsection{Inspiral}

Though the most common star in a typical star forming region is 
likely to be a low-mass dwarf, such objects are not likely to be
the most common partners in merger.  Low mass objects have small 
gravitational radii and thus interact effectively only with a
small part of the disk or envelope surrounding a massive collision
partner.  Additionally, mass segregation and competitive accretion 
in a dense cluster-forming cloud core are likely to lead to a 
relatively large median mass in the region where mergers are possible.
Thus, the most common interactions leading to a merger are likely
to involve relatively massive stars.  The most massive stars in the 
densest proto-clusters may experience multiple merging events.

A typical merger is expected to start with the dissipative passage 
of a star through the outer parts of a massive star's circumstellar
disk.  The order of magnitude of the density and column density of the
circumstellar debris field can be estimated by assuming that
an initial disk with a mass $M_d = 0.1$ \msol\ is ejected into
a spherical volume.   If ejected with a speed characterized by a fraction
of the orbital speed at the impactor periastron distance -- say
10 to 20 km/s corresponding to one quarter to one half of the Kepler
speed at 10 AU from a 20 \msol\ star, the outer edge of the debris
field would reach a radius of 1000 AU in 200 -- 400 years.  At an
outer radius of 1000 AU, the volume-averaged density of the debris
produced by the disruption of a 0.1 \msol\ disk is
$n(H_2) \sim 2 \times 10^6$ cm$^{-3}$,
and its column density is order $N(H_2) \sim 5 \times 10^{22}$ cm$^{-2}$.
The density distribution will probably be very inhomogeneous.
The encounter may form an eccentric binary in which one or both stars 
are surrounded by truncated disks.

Merging requires orbital decay.  As discussed above, accretion from
a surrounding cloud adds mass to the system, shrinks the orbits of
binary companions, and forces further dissipative interactions with 
circumstellar material.   The resulting disk perturbations during 
periastron can increase accretion form the disk onto its central star, 
temporarily increasing its luminosity.    In high density proto-clusters, 
interactions of single (or binary) stars with binary systems can also
also lead to merging (Bonnell \& Bate 2002). 

\subsubsection{\it Merging} 

The actual merger is likely to be a very short-lived event with
a duration comparable to the orbital time-scale at the tidal
radius.  Stellar collisions have been modeled by several groups
(e.g. Lombardi et al. 2003;  Freitag \& Benz 2004).
The merger parameters are likely to be similar to the case of the 
collision between two main sequence stars at a high impact parameter
where the velocity-at-infinity is small compared to the escape speed
from the stellar surface (e.g. Figures 23 and 24 in Freitag \& Benz, 2004).    
Most of the energy of a merger is released during this short phase.
While the collision product will cool and relax to a new main-sequence
configuration on a Kelvin-Helmholtz time-scale as discussed above,
the energy injected into the surrounding circumstellar debris
can emerge on a much shorter time-scale, especially if this debris
has a roughly disk-like geometry.

Most of the merger energy is likely to be radiated by the outer surface
of the merger envelope at infrared and sub-mm wavelengths. The observable
luminosity peak is likely to last for an envelope cooling time
which can range from years for small envelopes to centuries or more 
for massive envelopes.  There may be additional reprocessing of this
radiation by dense gas and dust surrounding the collision product. 

\subsubsection{\it Outflow} 

Observations show that accretion tends to be accompanied by outflow, 
typically with a speed comparable to the gravitational escape speed 
from the launch region (for reviews, see Reipurth \& Bally 2001;
K\"onigl \& Pudritz 2000; Shu et al. 2000).  
Thus, the rapid release of gravitational potential 
energy by a merger is likely to drive a powerful outflow into the 
surrounding medium.  The acceleration of such a flow must involve 
radiation fields and/or magnetic fields.   

If outflow of mass $M_{out}$ were to be accelerated only by 
mechanical means,  energy conservation implies that that the 
outflow kinetic energy at infinity $E_{out}$ 
must be less than the potential energy released by the mass accreted, 
or $E_{out} < G M (m - M_{out}) / R$.  Thus, the kinetic energy
of an outflow powered only by mass motion can at most
be $E_{KE} = 0.5 m v_{clus}^2$ $\approx 10^{45} m_1 v_{10}^2$ erg.

The energy produced by a merger may be transported by radiation
or magnetic fields from where it is released deep in the potential
well of the collision product to material at larger distances.  
In this case, a significant fraction of the gravitational binding energy
released by mass accretion onto the collision product can be coupled
to matter far away where it does not need to climb out of the
gravitational potential well.

The high luminosities of mergers involving at least one massive
star, combined with the large opacity of circumstellar debris,
are likely to make radiation effects an important ingredient 
for merger induced outflow acceleration.  Assuming that a typical 
debris disk surrounding a collision product contains 1 \msol\  
and is 100 AU in radius, its (spherical) average column density 
as seen from its center is $N(H_2) > 5 \times 10^{25}$ cm$^{-2}$ and 
$n(H_2) \approx 3 \times 10^{10}$ cm$^{-3}$.  Thus the circumstellar 
environment of the collision product will be opaque in most directions. 

An absorbing layer with a column density of order $10^{21}$ 
cm$^{-2}$ at the inner edge of this envelope will be compressed 
and accelerated by the radiation field.  The accelerated layer will
slam into gas in the envelope, driving a shock wave away from the 
collision product.   The radial velocity of this layer can be estimated
from the balance between radiation pressure and the ram pressure
of the advancing shock along each radial direction from the merger
product.  If the pre-shock mass density along each radial direction 
at a distance $r$ from the merger remnant is $\rho (r)$,  the 
radiatively accelerated shell along that direction will have a velocity 
$$
v_{shell}(r) 
  = \Bigl[ { {L} \over { 4 \pi c \rho (r)}   } \Bigr]^{1/2} { 1 \over r}
\approx 32.5 \Bigl[ { {L}      \over {10^5  L_{\odot}   }} \Bigr]^{1/2}
	    \Bigl[ { {n(H_2)} \over {10^7 {\rm cm^{-3}} }} \Bigr]^{-1/2}
	    \Bigl[ { {r}      \over {100 AU}             } \Bigr]^{-1} 
	    {\rm ~~~ (km / s^{-1})}
$$
This debris is likely to be flattened with a higher density in its 
equatorial plane than along its poles.  Thus, the radiatively driven shell 
may develop into a roughly bipolar pattern.   
In the dense disk, the radiatively driven layer will stall with a 
speeds of less than 1 \kms .    However for a merger product luminosity of 
$L > 10^5$ \lsol , a density less than $10^7$ cm$^{-3}$, and 
a radius of 100 AU, the shell speed will be greater than 30 \kms .
The radiatively accelerated shell will have a mass
$M_{shell} \approx 4 \pi r^2 \mu m_H N_{21}$ 
$\sim 6 \times 10^{-5} r_{100}^2$ \msol\
where $r_{100}$ is the shell radius in units of 100 AU.
Thus, a high-luminosity pulse of radiation may produce an impulsive bipolar 
outflow.  

The shocks driven into circumstellar debris will produce high-density 
compressed layers and possibly the products of ``hot core'' chemistry 
such as H$_2$O and SiO in the shock-heated gas.  Infrared radiation 
from grains heated by shocks or the radiation field of the collision 
product may also excite maser emission.

Magnetic acceleration of an outflow is likely to be considerably more
efficient than radiative acceleration.
Massive star forming molecular cloud cores are thought to be 
threaded by relatively strong magnetic fields in excess of 100 $\mu$G
(e.g. Lai et al 2003).  UV radiation from stars and shocks likely provide 
the minimum level of ionization required to couple these fields strongly 
to the entangled medium.  Fields threading the individual circumstellar
environments of merging protostars are likely to be entrained and
amplified by differential rotation and turbulent dynamo action during
a merger.  Expansion of the debris disk during the late phases of a 
merger may stretch entrained magnetic fields.  Thus, it is reasonable 
to expect that strong magnetic fields thread the debris 
surrounding a collision product.  Magnetic fields may convert a significant 
portion of the merger-released gravitational potential energy into the 
kinetic energy of an outflow by either the transient acceleration modes 
proposed by Uchida \& Shibata (1983; 1985), or the steadier disk-wind 
solutions of Pudritz \& Norman (1983).  

If the outflow is predominantly accelerated by the magneto-centrifugal
mechanism, it likely to start during the early phases of
inspiral.  As the magnetic field gets increasingly wrapped-up and the 
impactor spirals closer to the more massive star, the launch point will 
migrate deeper into the gravitational potential well of the system.  
Under the assumption that the flow terminal velocity is proportional 
to the escape speed from the wind launch point, the speed of the 
ejecta is likely to increase as the stars spiral towards each other,
reaching a maximum value as the merger completes.  Thus, an
eruptive, wide-angle outflow, with increasing mass loss rate
and speed that peak immediately after the merger is expected.  
As faster ejecta overtakes slower moving debris launched earlier 
during the inspiral phase, Rayleigh-Taylor instabilities are likely 
to develop (Stone, Xu, \& Lee 1995; McCaughrean \& Mac Low 1997).  
Such instabilities may produce the multiple fingers of high-velocity
shock-excited H$_2$ and [\Feii ] emission observed in Orion's OMC1
outflow (see below).

As the collision product relaxes, both stellar luminosity and outflow 
activity are expected to decline.  Thus, merger-generated outflows may
be impulsive when compared to the typical ages of protostellar
outflows from low to intermediate mass stars which range from 
$10^4$ to over $10^5$ years.  When observed just after the merger
event,  merger generated outflows may resemble explosions.  
After the eruption, but before the outflow has suffered significant 
deceleration by the surrounding medium, the ejecta may be characterized 
by a Hubble law with $v \propto d$.

\subsubsection{Disk Regeneration}

Two distinct mechanisms can lead to the re-formation of a 
circumstellar disk following a merger.
First, as the merging stars approach within their tidal radii,
the lower mass pre-main-sequence star will be tidally sheared 
(see the simulations shown in Zinnecker \& Bate 2002).  The 
resulting debris will form a torus surrounding the collision product.   
Merging of the two stars requires that orbital angular momentum be 
carried away.  Angular momentum can be transferred from the merging 
stellar cores to the outer portion of the debris surrounding the merger 
by gravitational torques.  Thus, as the stars spiral into each other, 
the outer edge of the surrounding debris disk must expand to larger radii.  
Some of this material may survive in orbit around the collision product 
as a disk and may serve as the dissipative medium for subsequent encounters.
Second, if the merger occurs inside a dense cloud core, continued infall 
may re-build a new post-merger accretion disk at a rate roughly given 
by $\dot M \sim c^3 / G$ where $c$ is the sound speed in the core.    
Disk regeneration by either or both mechanisms leads to the development 
of a dissipative circumstellar medium that sets the stage for additional 
mergers with any additional stars or binaries that wonder too close to 
the collision product.

The toroidal debris disk will partially collimate the outflow emerging
from the center into a roughly bipolar pattern.  Debris expanding into the
plane defined by the in-spiral orbit is likely to encounter the densest
material and suffer the greatest deceleration, while debris launched
along the axis of the system will encounter the least resistance.
The entire debris disk is likely to expand radially due to the combined
effects of dynamical heating, radiation pressure, and the impact of
fast ejecta from the core.

\section{Observable Consequences}

Protostellar merging events have several observable consequences
resulting from the release of gravitational potential energy
and the nature of the environment.  The energy released emerges
in one of three forms; radiation, kinetic energy of an expanding
high velocity outflow ejected mostly in the polar direction, and an
expanding lower velocity equatorial torus or debris disk.
Mergers are most probable in ultra-dense proto-clusters
and are likely to be highly obscured.

\subsection{Infrared Flares}

Heating of remnant circumstellar disks during the capture and
subsequent inspiral phases and the terminal merging event are 
expected to produce luminous flares in the thermal-infrared 
to sub-mm wavelength region that last an envelope cooling time 
which may range from years 
to millenia.  This prompt emission may be followed by a slower decline
in luminosity as the collision product relaxes on a Kelvin-Helmholtz 
time-scale and its radiation is reprocessed by the circumstellar 
medium.   These considerations suggest that massive stars produced
by merging may exhibit infrared light curves similar to FU Ori type
outbursts experienced by low-mass young stars.
An abrupt increase in luminosity, perhaps preceded by a series of
lower amplitude events triggered by pre-merger periastron
passages, will be followed by first a rapid decline with a time-scale
given by the envelope cooling time, followed by a much longer decline
to the collision product main-sequence luminosity on a 
Kelvin-Helmholtz time-scale.

Infrared flares should be most common in the densest cluster-forming 
regions in ultra-dense environments near galactic nuclei, in recently 
merged galaxies supporting high rates of star formation, and in 
regions producing super star clusters. 
Monitoring of the 5 $\mu$m to 1 mm wavelength fluxes produced by massive 
star forming regions in the Milky Way over several decades may provide 
one test of the merging hypothesis.   The event rate for infrared flares
ought to be comparable to the formation rate of massive stars by this
channel.   As discussed above,  the merging of two 100 \msol\ 
protostars can release over $10^{51}$ ergs, comparable to a supernova 
explosion.  If radiated on a $10^3$ year time-scale, this results 
in a flare with a luminosity larger than $10^{7}$ L$_{\odot}$.
However, such high luminosity events will be very rare 
compared to the lower luminosity events.

The rate of merger-produced flares should be proportional to the
merger rate.  If merging is an important pathway to the formation of 
massive stars, then it is expected that some high luminosity IRAS 
sources in our galaxy have brightened or faded since the IRAS observations
were made.  Monitoring of starburst regions in nearby galaxies such as 
M82 or NGC 253 may also reveal a class of time-variable infrared sources.  
Merger induced flares might be distinguished from other types of 
stellar variability by their high luminosity, long decay time, and 
absence of expanding supernova remnants.  Mergers should be highly 
confined point sources. Therefore, they would be best studied with 
large aperture telescopes such as the James Webb Space Telescope or 
with future 30--100 meter class ground based facilities equipped with 
advanced AO systems. 

In summary, mergers are expected to generate high luminosity infrared
flares at a rate comparable to the birth rate of massive stars 
by this mechanism.  Such flares may have rapid rise times, followed by
a relatively prompt decay lasting years to decades, followed by
slow decline to the main-sequence luminosity of the collision 
product on a Kelvin-Helmholtz time-scale.

\subsection{Impulsive Wide-Angle Outflows}

If 10\% of the energy released by the merger of a 1\msol\ star with
a 20 \msol\ star to form a merger product with a radius of 
$10^{12}$ cm is converted into bulk kinetic energy of motion of 
1 \msol\ of the surrounding medium, this medium can be accelerated to 
a terminal velocity of about 200 \kms .  
Outflows along the merger axis and merger
plane are likely to behave quite differently.  The portion of the outflow
that erupts orthogonal to the dense debris in the merger plane will
experience less resistance and is likely to be partially collimated by
the debris disk.  If so, collimation is likely to be very poor and
defined mostly by the geometry of the torus.  Each lobe of such a
mechanically collimated bipolar outflow (as opposed to magnetically
confined flows) is likely to subtend an opening angle of about a radian or more.
As the collision product relaxes, the merger-generated luminosity
and outflow activity will decline.  Thus, merger-generated outflows are 
expected to be dominated by the energy release occurring on a 
time-scale very short compared to the time required to form a star;  
such flows will resemble explosions.  

In summary, 
merger generated bipolar outflows are expected to be impulsive and
poorly collimated.  The flow velocity is expected to increase as the
merger progresses, perhaps triggering the formation of instabilities
in the ejecta.  

\subsection{Expanding Thick Disks}

A portion of the outflow energy will be coupled to the circum-merger torus.
The above energy estimates for the merger of 1 and 20 \msol\ protostars
indicate that much of a 1 to 10 \msol\ torus
can be accelerated away from the collision product to well above escape speed.
For example, if the thick torus intercepts 10\% of the total outflow energy
(corresponding to 1\% of the energy released by the merger) of a 1 \msol\
object merging with a 20 \msol\ star, a 10 \msol\ torus can be
accelerated to about 20 \kms .  The result would be an expanding field of
relatively low velocity debris in the merger plane.

Shocks driven into the circum-merger torus will produce very high-density
shock- and infrared-heated layers having conditions ideal for the excitation
of maser emission.  The resulting maser spots should trace the expansion
of the inner disk away from the forming
massive star.  Thus, the debris disk and outflows produced by mergers
may be sites of expanding clusters of OH, H$_2$O, SiO, and methanol
masers.  Orbital motion may shear these layers into arcs centered
around the central collision product.  These shocks may produce arcs-of
maser emission such as those observed in Cepheus A
(Torrelles et. al. 2001a,b).

In summary,
post-merger disks are expected to be dynamically heated thick
tori which may exhibit large expanding motions and contain shocks
which may be detectable as expanding arcs of maser spots.

\subsection{Uncorrelated Orientation of Multiple Eruptions}

Massive stars formed by a sequence of mergers with several 
lower mass stars may produce multiple widely-separated (in time) 
explosions of outflow activity with randomly oriented tori and 
outflow axes.  Thus, the outflow axis of a merger-grown massive 
star should meander chaotically over time.  Multiple uncorrelated 
(in orientation) outflow episodes may leave behind outflow cavities 
whose walls consist of filaments of dense gas pointing away from 
the collision product.  In OMC1, dozens of low-velocity NH$_3$ 
and dust filaments point radially away from the cloud core, perhaps 
tracing the walls of cavities left behind by successive periods of 
outflow activity (e.g. Wiseman \& Ho 1998; Johnstone \& Bally 1999).

In summary,
the formation of the most massive stars by protostellar mergers
may involve multiple merging events.  The orientations of the eruptive
outflows and expanding tori will exhibit a random walk in orientation.

\subsection{Decaying Compact Radio Continuum Sources}

The hot plasma generated by a protostellar merger may produce a
short-lived ultra-compact \hii\ region which decays on the time-scale
of the event.   Though the central portion of such a source will 
be sufficiently dense to be optically thick at all radio frequencies, 
plasma that escapes orthogonal to the debris disk, or which is 
entrained by the MHD outflow may become optically thin at high frequencies.

The combination of magnetic fields and strong shocks may lead to in-situ
particle acceleration.  Thus mergers may also excite some non-thermal radio
emission.  Some time-variable radio sources observed in nearby galactic nuclear
starbursts such as M82 and NGC 253 (Kronberg et al. 2000) may be powered by
proto-stellar mergers.   If a large fraction of massive stars form from
mergers, the merging rate could be comparable to the SN rate.  
The high-mass star formation rate in the nucleus of M82 appears to be about 
1 M$_{\odot} ~yr^{-1}$, leading to a supernova rate of about 1 every 
10 to 30 years, consistent with the number and decay rates of the observed 
time variable-radio sources.  It is also possible that fast shocks in the 
circumstellar environment of a merger may produce some X-ray emission.  
However, the expected large column density of this environment is should 
absorb most of this radiation.  Hard, penetrating radiation may however 
contribute to partial ionization of the medium which may lock associated 
magnetic fields to the medium.

In summary, protostellar mergers may produce relatively transient
ultra-compact thermal and possibly non-thermal radio sources that
decay on time-scales of years to centuries.

\subsection{Binaries and the IMF}

The origin of the initial mass function (IMF) has been extensively debated
in the literature (e.g. Larson 2005; Mac Low \& Klessen 2004; 
Padoan \& Nordlund 2002; Kroupa 2001, 2002; Elmegreen 2000; 
Bonnell et al. 1997, 2004).
While initial conditions in the cloud core and feedback in the
form of outflows and radiation may determine the masses of stars
formed in isolation, in clusters other factors may dominate.
Massive objects accrete at higher rates from a given environment than
less massive ones (competitive accretion).  Furthermore, objects near
the high density center of a core accrete faster than those in
lower density environments.  Protostellar merging provides an additional
mechanism which can alter the IMF (Bonnell, Bate, \& Zinnecker 1998;
Stahler et al. 2000).
Since a large fraction of field stars form in clusters that dissolve
soon after formation, these various processes are all likely to contribute
to the shape of the field IMF.   The more high-density clusters contribute to
the field-star population,  the larger the
role of mergers in determining the IMF.  Mergers may be the
dominant process in ultra-dense regions where systems such as 30 Doradus and
proto-globular clusters form.  Numerical simulations indicate that the 
expected mass-spectrum of massive stars is consistent with the predictions 
of the merger hypothesis (Bonnell \& Bate 2002).

An environment where the stellar density is sufficient to lead to
mergers is also expected to produce a large number of multiple star 
systems through three-body dynamical interactions and by means of 
captures of intruder stars that fail to merge (Bonnell \& Bate 2002; 
Bate, Bonnell, \& Bromm 2002).  Indeed, multiple systems
are more common among O and B stars than among lower mass objects
(Preibisch et al. 1999; Zinnecker \& Bate 2002).  A remarkable feature 
of the four massive stars in the Trapezium in Orion is that at least 
three of the four members ($\theta ^1$~Ori A, B, and C) are resolved 
multiples (Weigelt et al. 1999; Schertl et al. 2003).  Additionally, 
the brightest components of members A and B are also eclipsing binaries.  
Thus, the number of companions to the four massive Trapezium stars is 
unusually large; at sub-arcsecond scales, they consist of at least 
13 individual objects.

In summary, 
if merging dominates the birth of stars above a certain mass, then
one possible signature of this mass is the presence of a break in the
IMF at this mass, above which the IMF is steeper (e.g. Kroupa 2004).
Additionally, the multiple star fraction will increase above this
mass scale.

\subsection{Clustering and Run-away Stars}

Mergers can only occur in ultra-high density protostellar
environments with over $10^7$ YSOs per cubic parsec.
The observed densities of visible star clusters are never 
this high.  However,  such high densities are only expected
during the highly embedded phase of cluster formation. 
By the time most clusters become visible at visual or near-IR
wavelengths, they may have expanded significantly.  Deep
thermal-IR or high angular resolution radio wavelength 
surveys of dense cores such as the IRc2 region in Orion
are needed to determine the peak cluster density reached between
formation and re-expansion.  A rough estimate of
the peak cluster densities prior to proto-cluster
expansion can be made by assuming that such clusters form 
from the densest molecular cloud cores which are observed 
to have $\rm n(H_2) \approx 10^7 cm^{-3}$.
For a star formation efficiency
$\epsilon$, a median stellar mass $\bar M$, the density of a cluster formed
from a cloud with mean density $\rm n(H_2)$ is
$N_*$ = 
$2 \times 10^6  ~ \epsilon _{0.5} \bar M_{0.3}^{-1} n_7 $ $\rm stars ~pc^{-3}$
where $\bar M_{0.3}$ is the median stellar mass in units of 0.3 M$_{\odot}$,
$n_7$ is the density of molecular hydrogen in units of $\rm 10^7~ cm^{-3}$,
and $\epsilon _{0.5} = 0.5$ is the star formation efficiency (the ratio of
the total mass of stars formed, divided by the initial mass of gas).

Runaway high-velocity stars are thought to be produced by the 
break-up of binaries in which one member exploded (Blaauw 1991) or
from three- and four-body (single--binary and binary--binary) 
interactions.   The contribution of the latter process grows in 
importance with increasing cluster density.  The massive run-away 
stars AE Aurigae and $\mu$ Columbae apparently originated from a
binary--binary interaction which 2.6 Myr ago produced two fast 
run-away stars and the binary star $\iota$ Orionis located in 
the Orion OB1c subgroup in front of Orion's Trapezium cluster 
(Hoogerwerf et. al. 2000).   Only such dynamical interactions can 
produce high-velocity runaway stars {\it before} the first supernova 
(SN) explosion in a group (Kroupa 2004).  Clusters dense enough 
to promote merging will develop higher velocity dispersions than 
groups that do not allow it.  Thus, clusters which produce large 
numbers of high-velocity run-away stars and stars with velocities much 
larger than the escape speeds from cluster-forming molecular cloud 
cores provide indirect evidence for a prior epoch during which 
stellar densities were sufficiently high to promote dynamical 
interactions.  Lower cluster densities would result in fewer 
and slower runaway stars generated by three- and four-body 
interactions, few interaction-formed binaries, and fewer rapid 
rotators.

In summary, the statistics of run-away stars can be used as an indicator
of the high cluster densities required for merging to occur.  Deep, 
high angular resolution radio and far-IR wavelength studies can be used to
directly measure the stellar densities of embedded cluster during
their formation.

\subsection{Gamma-ray Bursters and Hypernovae}

Gamma-ray bursters (GRBs) are known to be associated with massive star 
forming regions in galaxies (see Meszaros 2002 for a review).  
The collapsar model for GRBs requires a rapidly rotating and very massive 
progenitor star (Woosley 1993; Proga et al. 2003).  The off-axis collisions 
and merging of stars near the top of the stellar mass spectrum may result 
in the production of GRBs.  The rare merger of two (or more) stars 
at the very top of the stellar mass spectrum are most likely to occur in 
the most extreme star formation environments such as those 
which produce super star-clusters in galactic starbursts (e.g. the Antennae)
and dwarf galaxies (e.g. 30 Dor in the LMC).  Such events can release
more than $10^{51}$ ergs of gravitatinal potential energy.

The merging of massive stars may produce super-massive ($>$ 100 \msol)
collision products with rapid rotation since the most probable collisions occur 
at high impact parameter.  Rapidly rotating super-massive stars are likely 
to be near the Eddington limit, radiate most of their luminosity at 
short ultraviolet wavelengths, and be hotter at their poles and darker 
at their equators.   Radiation, and the resulting line-driven 
stellar winds, will tend to emerge from the poles of the rotationally 
flattened object (von Zeipel 1924; Owocki et al. 1998; Smith et al. 2003).  
Thus, surrounding interstellar gas will be cleared most efficiently along the 
stellar rotation axis.  

The cores of massive stars are convective.  The combination of rapid spin 
and convection may lead to the generation of a strong internal magnetic field. 
The super-massive object may be eventually disrupted by an unusually 
violent supernova explosion (a hypernova) soon after formation.  
During the hypernova, the rapidly spinning core of the collapsing star 
may form a transient, magnetized toroid before its mass plunges into 
the black hole forming at its center.  As matter spirals toward and
into the black hole, some of its gravitational potential energy may be 
efficiently converted into relativistic jets launched by the magnetic
field along the spin axis.  These conditions may be ideal for the production 
of relativistic jets that blast through the remains of the star and into 
relatively clear regions along the object's spin axis.  As these jets slam 
into residual gas in the cavities along the merger product's rotational 
axis,  they may emit hard gamma and X-ray radiation.  Thus, the extremely 
rare merger of massive stars may produce gamma-ray bursters (GRBs).

The delay between the formation of a super-massive star and its death
in a hypernova explosion is likely to be comparable to the nuclear
time-scale for the star.  This delay can be estimated from
$\tau_{nuc} \approx 0.007 \epsilon M c^2 / L$  years where
$\epsilon$ is the fraction of the star involved in nuclear energy release 
before detonation.  For $\epsilon = 0.25$ (probably a severe upper limit), 
$M = 120$ \msol , and $L = 1 \times 10^6$ \lsol , $\tau_{nuc}$ is about 3 Myr. 

A very rough estimate of the rare merger rate of a pair of stars near
the top end of the IMF can be obtained either by considering the rate 
of bound cluster formation in the Milky Way or the formation rate
of the most massive stars.   The bound cluster birth rate has been
estimated to be in the range 0.1 to 0.4 kpc$^{-2}$ Myr$^{-1}$ in the
Solar vicinity (e.g. Lada \& Lada 2003).  Assuming that most such 
clusters form in the Molecular Ring which has an area of about 
170 kpc$^2$, about 17 to 70 such clusters form somewhere in the 
Milky Way every million years.  If 10\% of these bound clusters 
produce very massive stars near the upper mass limit of around 
120 \msol , the birth rate of such stars is around 1.7 to 7 Myr$^{-1}$.  
Such stars live only about 3 Myr.  Thus, there will be of order 10 
such stars in the Galaxy at any one time, consistent with the number 
known in the 100 \Msol\ range.   GRBs are thought to have a beaming 
angle of about 4\arcdeg\ (Frail et al. 2001) so that only about 0.1\% of 
all GRBs will be visible from Earth.  If all super-massive stars are merger 
remnants with fast rotation that produce hypernovae, and 0.1\% of the 
resulting GRBs are oriented towards us, then we would expect to see one Galactic 
GRB every 0.1 to 1 billion years.  If only 10\% of super-massive stars 
form by merging, the GRB rate would be 10 times lower.  However, the 
star formation rate may have been 10 times higher in the past 
(e.g. at a redshift of $\sim$ 1).  There are $10^{11}$ similar galaxies in 
a Hubble volume.  Under these assumptions, the total GRB rate is expected 
to be between $10^2$ to $10^3$ events per year, not inconsistent with 
the actual detection rate of GRBs.   

In summary, it is possible that the merging of the most-massive stars
lead to hypernovae and the production of one class of gamma-ray bursters.

\subsection{Distinguishing Mergers from Accretion}

Several types of observations can be used to distinguish between the
accretion and merger scenarios.  While accretion can in principle
produce massive stars in isolation, mergers can only occur in very
dense clusters that presumably can only form in very dense and massive
cloud cores.  Neglecting segregation of stars and gas and contraction, 
the formation of a cluster density of $10^6$ \msol\ pc$^{-3}$ requires a 
pre-star-formation cloud core gas density of $n(H_2) \approx 10^7 ~cm^{-3}$.
However, as shown by Bonnell et al. (2003), gravitational collapse of
a fragmenting core with an initial density of $n(H_2) = 10^5 ~cm^{-3}$ can 
lead to stellar densities of $10^6$ to $10^8$ stars per cubic pc for a brief
period. It is during this transient ultra-dense and highly embedded phase 
that mergers are most likely.  A corollary to the high density is that 
interactions will truncate disks.  In the Bonnell et al. (2003) simulations,
one-third of all stars and virtually all massive stars, suffer interactions
with periastron separations of less than 100 AU.

Circumstellar disks are expected to have very different properties in the
merger and accretion scenarios.  Accretional growth of a protostar 
is expected to be associated with the presence of quiescent, 
geometrically thin, accretion disks.  Close-encounters and mergers 
will disrupt such disks.  One expected signature of a recent merger 
is the presence of a dynamically heated, expanding torus of debris 
surrounding the collision product.  Dynamical stirring
and the increasing rate of energy dissipation as a system evolves towards
a merger may launch shocks into this expanding debris field.  The dense
and hot environment of a merger may be ideal for the production of maser
emission.  The merger and subsequent reprocessing of radiation to longer
wavelengths by the circumstellar debris may produce a high luminosity 
flare at far infrared wavelengths that lasts from years to centuries.
In the merger scenario, some encounters will lead to the formation
of captured binaries and when the merger is completed,
the resulting product will have rapid spin.  Thus, a high multiplicity 
fraction and fast rotation among massive stars may be one signature 
of interactions and merging.  Large numbers of high-velocity 
($>$ 50 \kms ) as well as moderate-velocity (10 to 50 \kms ) run-away
stars originating in OB associations and young star clusters can provide 
indirect evidence for extremely high stellar densities during their birth.

Outflows produce one of the most easy to observe signs of the birth of a
young star.  Disk accretion in forming low-mass stars produces
highly collimated jets (e.g. Reipurth \& Bally 2001).  Such jets have
been observed to be produced by moderate mass protostars with luminosities
larger than $10^4$ \lsol .  However, as discussed below, the outflow
from the nearest $10^5$ \lsol\ class protostar, OMC1, has a very different
morphology indicating an explosive origin.  While massive stars growing by
accretion from a disk may produce collimated outflows which are quasi 
scaled-up versions of those driven by low-mass protostars, mergers are expected
to drive wide-angle, eruptive outflows.  The formation of the most massive
stars may involve multiple protostellar merging events.  The orbital planes
defined by a series of in-spirals are likely to be random.
Thus, the orientations of the outbursts produced by a series of merging
events are likely to be un-correlated with each other.  The series of
eruptions would likely leave behind a set of outflow lobes and cavities
which point radially away form the massive collision product.
These different expectations of accreting and merging protostars are
summarized in Table~1.

\section{The OMC1 Outflow: Evidence for a Recent Merger?}

The Orion region contains the nearest region of on-going massive
star formation (e.g. O'Dell 2001).  The Trapezium cluster associated
with the Orion nebula, one of the youngest clusters near the Sun,
contains over 2000 mostly low-mass young stars less than about $10^6$
years old (Hillenbrand 1997; Hillenbrand \& Hartmann 1998)
The density of stars in the cluster core currently is in the range 
$5 \times 10^4$ stars pc$^{-3}$ (McCaughrean and Stauffer 1994) 
to $2 \times 10^5$ stars pc$^{-3}$ (Henney \& Arthur 1998),
implying a nearest-neighbor distance between low-mass 
stars of only a few thousand AU (e.g. Bally et al. 1998).

Over 50\% of the young stars in the Trapezium cluster are surrounded
by disks with radii up to 500 AU.  The intense UV radiation field of the
Trapezium stars is photo-evaporating these disks, making them
visible as Orion's `proplyds' (O'Dell, Wen, \& Hu 1993; O'Dell \& Wong 1996;
Bally et al. 1998, 2000; Johnstone, Hollenbach, \& Bally 1998; McCaughrean, 
Stapelfeldt, \& Close 2000).  In this cluster, the present ratio of the 
average interstellar separation to the typical disk radius is only 10 to 
100.  Since these disks are losing mass at rates around $10^{-7}$ to 
$10^{-6}$ \msol\ per year, they were more massive in the recent past.  
Since star formation in the Orion nebula is finished (except in dense 
cores such as the BN/KL region behind the nebula) the visible portion 
of the cluster is likely to be in a state of expansion as the surrounding 
molecular cloud that dominates the mass in the region is photo-ablated.   
Thus, when the massive Trapezium stars formed,  the cluster was probably 
denser, and its low-mass stars had more massive disks.  A 3 \kms\ velocity 
dispersion for the Trapezium cluster and a core radius of 0.1 pc implies 
a crossing-time of about $3 \times 10^4$ years.  So, this cluster could 
have expanded significantly in the recent past.  These factors make 
it plausible that mutual interactions and cannibalism might have occurred 
prior to the emergence of the massive Trapezium stars.  Indeed, the large 
multiplicity of the massive Trapezium stars and the gap in the radial 
distribution of low-mass stars surrounding $\theta ^1$~Ori~C (Hillenbrand 1997)
may be evidence that this star has merged with its neighbors while it 
was forming.

Massive star formation is still occurring within the luminous
($10^5$ \lsol ) OMC1 core located immediately behind the Orion nebula.
Though many individual peaks of thermal infrared (7 to 24 $\mu$m) emission
have been identified (Lonsdale et al. 1982), many may not be self-luminous.
The VLA studies of Menten \& Reid (1995) indicate that, in addition to
the Becklin-Neugebauer object (BN), the region contains at least two
other ultra-compact radio sources, I and n, separated
by about 3\arcsec (1,500 AU in projection) from each other.

Radio sources I and n in OMC1 are both associated with strong 
OH, H$_2$O, and SiO maser emission (Johnston et al. 1989; 
Genzel et al. 1981; Greenhill et al. 2003; 2004). 
An expanding arcminute-scale complex of high velocity ($v$ =
30 to 100 \kms ) H$_2$O masers surround the entire  OMC1 region.
The expansion center coincides within several arcseconds of source n
(Genzel et al. 1981). In addition to this high velocity flow, a
low velocity ($v$ = 18$\pm$2 \kms ) cluster of much brighter maser
spots is associated with this expansion.  As shown by
Gaume et al. (1998), the so-called 22 GHz H$_2$O `shell'
masers, which were mostly resolved-out in the VLBI observations
of Genzel et al. (1981), are concentrated in a 2\arcsec\
by 0.5\arcsec\ strip centered on source I and oriented roughly
orthogonal to the bipolar CO outflow emerging from OMC1 
(Chernin \& Wright 1996).  Additionally, bright SiO maser emission 
lies within 0.1\arcsec\ of source I.  These masers consist of 
four clusters of emission having velocities very similar to 
those of the H$_2$O shell masers.  Greenhill et al. (1998) and 
Doeleman, Lonsdale, \& Pelkey (1999) found that these masers are 
concentrated into four linear chains fanning out to the North, 
East, South, and West from source I.  The North and West clusters 
are redshifted while the south and east clusters are blueshifted.  
Thus, these SiO maser chains have Doppler shifts opposite to the 
CO flow emerging from this region.

A fast (30 to 100 \kms), poorly collimated bipolar outflow
emerges from this region orthogonal to the `shell' maser disk
with a blueshifted lobe towards the northwest.  The OMC1 outflow
has a mass of about 10 \msol\ and a kinetic energy of about
$4 \times 10^{47}$ ergs (Kwan \& Scoville 1976). The OMC1 outflow
and the H$_2$ fingers (Allen \& Burton 1993; Stolovy et al. 1998;
Schultz et al. 1999; Kaifu et al. 2000) indicate that
a powerful explosion occurred in OMC1.  The H$_2$ finger system 
consists of over a hundred individual bow-shocks
which delineate a relatively wide-angle bipolar outflow
towards PA $\approx$ 315\arcdeg\ with with an opening angle
of more than 1 radian in each lobe.  Some of the H$_2$ fingers are
visible in Hubble Space Telescope images and thus have known
proper motions.  The proper motion vector field (Lee \& Burton 2000;
Doi, O'Dell,\& Hartigan 2002) indicates an explosive origin about
1010$\pm$140 years ago, assuming no deceleration.  Greenhill et al.
(1998) interpreted the `shell' water masers and the 18 \kms\ outflow
as tracers of an expanding disk surrounding source I with a northeast--
southwest major axis that lies orthogonal to the fast wide-angle 
bipolar outflow that emerges along a northwest--southeast axis.
However, Greenhill et al. (2003) presented an argument, based on
new VLBA and 7 mm VLA data, that the disk is oriented southeast--northwest
and that the ``shell'' masers trace a jet.  In this picture, the 
four SiO maser chains trace the surface of the disk, which produces
a ridge of 7 mm radio continuum emission along the disk plane.  
However, this new re-interpretation leaves the H$_2$ fingers and
associated CO flow without a known driving source.  Furthermore, it
is possible that the 7 mm radio emission actually traces a
thermal radio jet rather than dust emission from a disk.
Thus, we believe the interpretation in which the ``shell'' and 
SiO masers surrounding source I trace a thick, expanding disk, 
and the 7 mm emission traces a jet oriented northwest-southeast 
which drives the H$_2$ fingers and associated CO outflow provides a 
simpler explanation for the various phenomena in OMC1.

The kinematics, energetics, and morphology
of the OMC1 region can be explained by a recent merger
suffered by source I.
A 20 \Msol\ star swallowing a 1 \Msol\ object will release about
$3 \times 10^{48}$ ergs of gravitational potential energy, more
than enough to drive an outflow such as that observed in the OMC1
cloud core.  Merger-generated outflows are expected to be
highly impulsive, poorly collimated, and may exhibit an ejection
speed that increases with time.
Such accelerating flows may readily produce the instabilities that
are required to explain the multiple fingers of H$_2$ emission in
Orion (Stone et al. 1995; McCaughrean \& Mac Low 1997).

Plambeck et al. (1995) and Tan (2004) found that the BN star,
located about 10\arcsec\ northwest of the OMC1 core, 
has a large proper motion of 38.7 $\pm$ 4.7 \kms\ towards 
P.A. --37.7 $\pm$ 5\arcdeg .  Additionally, Scoville et al. (1983)
found that this source is moving about +12 \kms\ with respect to
the OMC1 cloud core in the radial direction.  Tan (2004) interpreted 
this motion as the result of ejection from the Trapezium some 
4,000 years ago.  He notes that BN passed within 1\arcsec\ of
source I about 500 years ago.    

One problem with the interpretation that BN is a runaway star 
expelled from the Trapezium is that it exhibits traits of a star 
much younger than any of the Trapezium members.   While none of
the Trapezium stars are surrounded by obvious circumstellar
environments, BN is surrounded by an envelope that produces 
600 -- 900 K dust continuum and absorption/absorption in the 
near-infrared vibrational bands of CO (Scoville et al. (1983).  
Hot (3,500 K) CO band-head emission emission is formed by a dense 
($n(H_2$ = $10^{11}$ to $10^{13}$ cm$^{-3}$) molecular layer having 
a very small ($<$0.7 AU) area.   Scoville et al. (1983) suggested that
this emission may trace a shock front, or possibly a disk surrounding 
a low-mass companion. In addition to a large infrared excess, and 
infrared emission from CO, BN is surrounded by an ultra-compact 
($\sim$ 20 AU radius) \Hii\ region, and exhibits Brackett-series 
emission lines  that have been interpreted as a decelerating 
stellar wind with a mass loss rate of 
$\dot M \approx 4 \times 10^{-7}$ to
$10^{-7}$ \msol yr$^{-1}$ and a relatively low terminal
velocity of less than a hundred \kms .  Furthermore, 
redshifted absorption in CO indicates infall onto BN from 
beyond 25 AU at rates of 
$\dot M \approx 4 \times 10^{-6}$ \msol yr$^{-1}$. 
Thus, this ejected high-velocity star must be dragging along 
a substantial ($> ~ 10^{-4}$ \msol ) reservoir of dense circumstellar 
gas.  If BN originated from the Trapezium, then it is puzzling 
why it is surrounded by such a rich circumstellar environment 
when none of the Trapezium stars exhibit such media or other 
youthful traits.    

We consider the possibility that BN was expelled not from
the Trapezium, but from the OMC1 cloud core, possibly by 
source I.   If the outer parts of the OMC1 outflow for which
proper motions have been measured has been decelerated by a 
factor of about two,  then the true age of the outflow is 
about 500 years.  Proper motions indicate that BN passed within 
1\arcsec\ of source I about 500 years ago.  It would be a 
remarkable coincidence for a high velocity star to pass 
so close to another massive star, and to have this occur just 
when a remarkable outflow is launched.  A physical interaction 
with source I might explain the OMC1 outflow, the ejection of BN,  
and by its moving through the gas-rich environment of the OMC1 
cloud core, the presence of a co-moving circumstellar 
environment.    

If BN was ejected from the vicinity of source I,  its 
+12 \kms\ radial velocity with respect to the stationary CO 
emission from OMC1 and proper motion towards the northwest 
would place it behind the northwest lobe of the OMC1 outflow.   
Indeed,  Scoville et al. (1983) found CO absorption at the 
velocity of the blueshifted lobe of the OMC1 outflow 
towards BN.  The ejection of BN requires the involvement of 
a third star.   This star either survives as a close companion of 
source I, or may have spiraled into and merged with this star.   
Either interaction may have triggered the ``explosion'' responsible 
for the powerful OMC1 outflow and associated H$_2$ fingers.

The NH$_3$ and dust filaments in the OMC1 cloud core
(e.g.  Wiseman \& Ho 1998; Johnstone \& Bally 1999)
point radially away from the vicinity of the BNKL core. 
These features may trace the walls of cavities left behind 
by earlier episodes of outflow activity in OMC1.
Some of these events may have been triggered by earlier 
protostellar interactions.

\section{Conclusions and Summary}

Stellar merging is possible in forming clusters during a very
short-lived and transient phase during which cluster densities
exceed $10^7$ stars pc$^{-3}$.  On the other hand, it is possible 
that in low stellar density environments or in isolated small cloud 
cores, stars of most masses form by direct accretion without 
the perturbations caused by sibling protostars.  Thus, there may 
be multiple pathways by which moderate to high-mass stars form.  
High resolution infrared and radio wavelength observations of 
forming star clusters are needed to determine the significance of 
mutual interactions and mergers.

Merging rates can be greatly enhanced
by gravitational focusing in low velocity dispersion
clusters and by the presence of circumstellar material in the
form of envelopes and disks which can provide
a dissipative medium.  Protostellar mergers can release between
$10^{45}$ ergs for the most common low-mass events to over $10^{51}$
ergs when a pair of 100 \msol\ stars merge.

The formation of massive stars by proto-stellar mergers can be
distinguished from the standard scenario of disk accretion 
by several types of observation.   Mergers are expected to produce 
unique phenomena including:

\noindent 1]
Luminous infrared flares with rise-times comparable to the orbital time-scale
of the merging protostars, followed by a fast decline dominated by the cooling 
time of circumstellar debris, and the longer lasting decline on the 
Kelvin-Helmholtz time-scale of the merger product.

\noindent 2]
Eruptive, wide-angle bipolar outflows which may be subject to
instabilities resulting from the collision of accelerating winds
powered by the increasing energy-release of the merger.

\noindent 3]
Expanding thick tori of debris produced by the merging of
circumstellar disks and the tidal disruption of the lower-mass
collision partner.  Shocks driven into this debris disk by the 
in-spiraling protostars may drive `hot core' chemistry, excite maser 
emission, and non-thermal radio emission.

\noindent 4]
Multiple merging and mass-loss episodes producing uncorrelated outflow
lobe orientations.   Such multiple eruptions may leave behind
radial fingers of swept-up gas and dust pointing away from the 
merger product.

\noindent 5]
Capture-formed binaries due to incomplete or failed mergers.  Merging 
is expected to deplete the immediate vicinity of a forming massive star 
of lower mass sibling stars.

\noindent 6]
Clusters with sufficient density to allow mergers may produce high
stellar multiplicity, stars with fast rotation,  and large numbers of
high-velocity run-away stars.

\noindent 7]
It is possible that the merging a pair of stars close to the top 
end of the IMF produce hypernova explosions which lead to gamma-ray bursts.

\noindent 8]
The BNKL outflow in the OMC1 cloud core behind the Orion nebula
is interpreted in terms of a merger model.  Several observed 
properties of this source fit the expectations of the merger model.

High resolution observations of infrared to radio-wavelength emission
from active sites of massive star formation are needed to distinguish between
the direct accretion and merger scenarios.  Studies of the disks, 
outflows, kinematics, and infrared photometric variability of high 
luminosity protostellar sources in the Milky Way will determine if mergers 
play a role in massive star formation.   
The morphology of shock-excited H$_2$ and [Fe II] emission in the 
near-infrared, fine-structure cooling lines in the mid-infrared, 
molecular tracers in the sub-millimeter and millimeter parts of the spectrum
will be used to trace outflow and disk structure.  Such observations will 
be feasible in the near future with Gemini, VLT, Keck, JWST, and aperture 
synthesis instruments such as ALMA and CARMA.  Monitoring of the 
thermal infrared and radio emission from point sources in massive star 
forming regions in starburst regions will provide an additional test of 
massive star formation models.  Such observations and determinations of 
stellar densities in highly obscured cloud cores may become feasible at 
thermal-infrared imaging with JWST, and at radio and sub-mm wavelengths 
with ALMA and E-VLA.  In the far future,  adaptive-optics assisted near 
infrared imaging with 30 to 100 meter telescopes may be able to penetrate the
extremely large (A$_V > $ 100, or A$_K$ $>$ 10) extinction of
cluster-forming cores as distance of a few kpc.  

Though the standard accretion scenario may well explain the formation of 
most massive stars, the existence of blue-stragglers in globular
clusters indicates that mergers do occur in nature. The formation phases 
of massive stars and clusters are still hidden from us by high extinction 
and source confusion.  Thus, merging in protostellar environments is a 
plausible pathway to at least some massive stars.  It is important 
to explore possible observable consequences that may distinguish the merging 
from the accretion scenarios.   This paper is intended to stimulate further 
work on the various possibilities.

\noindent
{\bf Acknowledgements:}
{We thank 
Mark Freitag, 
Ralf Klessen, 
Hal Levison, 
Bo Reipurth,
Nathan Smith, 
and 
Josh Walawender 
for useful discussions and new insights.   We thank the referee, Ian
Bonnell for many helpful comments.  This work was supported in
part by NASA grants NAG5-8108 (LTSA) and NCC2-1052 (Astrobiology),
and in part by the National Science Foundation under Grant No. 9819820.}

% REFERENCES

\section{References}

\noindent
Allen, D.~A. \& Burton, M. G. 1993, Nature, 363, 54

\noindent
Bally , J.  2002,
in {\it Hot Star Workshop III: The Earliest Phases of Massive Starbirth}

ed. P. Crowther, ASP Conf.Ser, 267, p. 219

\noindent
Bally, J., Sutherland, R. S., Devine, D., \& Johnstone, D. 1998, AJ, 116, 293

\noindent
Bally, J., O'Dell, C. R., \& McCaughrean. M.
2000, AJ, 119, 2919

\noindent
Bate, M, Bonnell, I., \& Bromm, V. 2002, MNRAS, 336, 705

\noindent
Bate, M, Bonnell, I., \& Bromm, V. 2003, MNRAS, 339, 577

\noindent
Beech, M., \& Mitalas, R.
1994, ApJS, 95, 517
% ``Formation and evolution of massive stars''

\noindent
Behrend, R., \& Maeder, A.
2001, A\&A, 373, 190
% ``Formation of massive stars by growing accretion rate''

\noindent
Benz, W., \& Hills, J. G. 1992, ApJ, 389, 546
% 3D collisions, unequal masses

\noindent
Bernasconi, P. A., \& Maeder, A.
1996, A\&A, 307, 829
% `About the absence of a proper zero age main sequence for massive stars''

\noindent
Bertoldi, F.
1989, \apj, 346, 735

\noindent
Beuther, H., Schilke, 
P., Sridharan, T.~K., Menten, K.~M., Walmsley, C.~M., \& 

Wyrowski, F.\ 2002, \aap, 383, 892 

\noindent
Blaauw, A.\ 1991, NATO ASIC 
Proc.~342: The Physics of Star Formation and Early Stellar 

Evolution, 125 

\noindent
Boffin, H. M. J., Watkins, S. J., Bhattal, A. S., Francis, N., \&
Whitworth, A. P.  

1998, MNRAS, 300, 1189

\noindent
Bonnell, I. A., \& Bate, M. R. 2002, MNRAS, 336, 659

\noindent
Bonnell, I. A., Bate, M. R., \& Zinnecker, H. 1998, MNRAS, 298, 93

%\noindent
% Bonnell, I. A. \& Clarke, C. J. 1999, \mnras, 309, 461

\noindent
Bonnell, I. A., Bate, M. R., Clarke, C. J., \& Pringle, J. E.\
1997, MNRAS, 285, 201

\noindent
Bonnell, I. A., Bate, M. R., Clarke, C. J., \& Pringle, J. E.\
2001a, MNRAS, 323, 785 

\noindent
Bonnell, I. A., Clarke, C. J., Bate, M. R., \& Pringle, J. E.\
2001b, MNRAS, 324, 573 

\noindent
Bonnell, I. A. 2002,
in {\it Hot Star Workshop III: The Earliest Phases of Massive Starbirth}

ed. P. Crowther, ASP Conf.Ser, 267, p. 193

\noindent
Bonnell, I. A., Bate, M. R., \& Vine, S. G. 
2003, MNRAS, 343, 413 
% Hierarchical formation of a stellar cluster

\noindent
Bonnell, I. A., Vine, S. G., \& Bate, M. R.
2004, MNRAS, 349, 735
% Oriing of the IMF

\noindent
Chernin, L. M., \& Wright, M. C. H. 1996, ApJ, 467, 676

\noindent
Chini, R., Nielbock, M., \& Beck, R.\
2000, A\&A, 357, L33
% ``The birth of massive twins in M 17''

\noindent
Chini, R., Hoffmeister, 
V., Kimeswenger, S., Nielbock, M., N{\" u}rnberger, D., 

Schmidtobreick, L., \& Sterzik, M.\ 2004, Nature, 429, 155 

\noindent
Clarke, C.~J.~\& Pringle, J.~E.\ 1991, \mnras, 249, 587
%Star--disk interactions in binary star formation

\noindent
Clarke, C.~J.~\& Pringle, J.~E.\ 1993, \mnras, 261, 190
% "Accretion disc response to a stellar fly-by"

\noindent
Clarke, C.~J.~\& Pringle, J.~E.\ 1992, \mnras, 255, 423

\noindent
Claussen, M. J., Gaume, R. A., Johnston, K. J., \& Wilson, T. L.
1994, ApJ, 424, L41

\noindent
DeWit, J. W., Testi, L., Palla, F., \& Zinnecker, H., 2004, submitted to A\&A

\noindent
Doeleman, S.~S., Lonsdale, C.~J., \& Pelkey, S.\ 1999, ApJ, 510, L55

\noindent
Doi, T., O'Dell, C. R.,  \& Hartigan, P. 2002,
AJ, 124, 445

\noindent
Elmegreen, B. G. 2000, ApJ, 539, 342

\noindent
Elmegreen, B. G.\ 2001, ASP Conf.~Ser.~243:
{\it From Darkness to Light: Origin and Evolution

  of Young Stellar Clusters}, 255

\noindent
Elmegreen, B. G. \& Scalo, J.  2004, ARA\&A, 42, 211
% Interstellar Turbulence I: Observations and Processes

\noindent
Frail, D.~A., et al.\ 2001, ApJ, 562, L55 

\noindent
Fregeau, J. M., Cheung, P., Portegies Zwart, S. F., \& Rasio, F. A.
2004, MNRAS, 

 (in press -- astro-ph0401004)

\noindent
Freitag, M. \& Benz, W. 2004, MNRAS (in press - astroph 0403621)

\noindent
Gaume, R. A., Wilson, T. L., Vrba, F. J., Johnston, K. J.,
\& Schmidt-Burgk, J.

 1998, ApJ, 493, 940

\noindent
Genzel, R., Reid, M. J., Moran, J. M., \& Downes, D. 1981,
ApJ, 244, 884

\noindent
Gies, D. R. 1987, ApJS, 64, 545

\noindent
Greenhill, L.J., Gwinn, C. R., Schwartz, C., Moran, J. M.,
\& Diamond, P. J. 1998, 

Nature, 396, 650

\noindent
Greenhill, L.J. et al. 2003, IAUS 221, 203 (Sydney)

\noindent
Greenhill, L.J. et al. 2004, ApJ 605, L57

\noindent
Hall, S.~M., Clarke, C.~J., \& Pringle, J.~E.\ 1996, \mnras, 278, 303
% "Energetics of star-disc encounters in the non-linear regime"

\noindent
Hanson, M.~M., Howarth, I.~D., \& Conti, P.~S.\
1997, \apj, 489, 698
% ``The Young Massive Stellar Objects of M17''

\noindent
Heller, C.~H.\ 1995, \apj, 455, 252
% "Encounters with Protostellar Disks. II. Disruption and Binary Formation"

\noindent
Henney, W. J., \& Arthur, S. J.  1998, AJ, 116, 322

\noindent
Henning, T., Schreyer, K., Laundhardt, R., \& Burkert, A.
2000, A\&A, 353, 211
% ``Massive young stellar objects with molecular outflows''

\noindent
Hillenbrand, L. A. 1997, AJ, 113, 1733

\noindent
Hillenbrand, L. A. \& Hartmann, L. W. 1998, ApJ, 492, 540

\noindent
Ho, P.~T.~P., Klein, R.~I., \& Haschick, A.~D.\
1986, ApJ, 305, 714
% ``Formation of OB clusters - Radiation-driven implosion?''

\noindent
Hoogerwerf, R., de Bruijne, J.~H.~J., \& de Zeeuw, P.~T.\ 2000, \apjl, 544,
L133

\noindent
Jijina, J. \& Adams, F. C. 1996, ApJ, 462, 847

\noindent
Johnston, K. J., Migenes, V., \& Norris, R. P. 1989, ApJ, 341, 847

\noindent
Johnstone, D. \& Bally, J. 1999, ApJ, 510, L49

\noindent
Johnstone, D., Hollenbach, D., \& Bally, J.
1998, ApJ, 499, 758

\noindent
Kahn, F. D. 1974, A\&A 37, 149

\noindent
Kaifu, N. et al. 2000, PASJ, 52, 1

\noindent
Kessel-Deynet, O. \& Burkert, A. 2003, MNRAS 338, 545

\noindent
Keto, E.
2002, ApJ, 568, 754
% ``An Ionized Accretion Flow in the Ultracompact H II Region G10.6-0.4''

\noindent
Klessen, R. 2001, ApJ 556, 837

\noindent
Klein, R.~I., Sandford, M.~T., \& Whitaker, R.~W.\
1983, ApJ, 271, L69
% ``Star formation within OB subgroups - Implosion by multiple sources''

\noindent
K\"onigl, A., \& Pudritz, R.~E.\ 2000, Protostars and Planets IV, 

eds. V. Mannings, A. Boss, \& A. Russell p. 759 

\noindent
Kronberg, P. P., Sramek, R. A., Birk, G. T., Dufton, Q. W.,
Clarke, T. E., \& Allen, M. L.  

  2000, ApJ, 535, 706

\noindent
Kroupa, P. 1995, MNRAS 277, 1522

\noindent
Kroupa, P. 2002, Science, 295, 82

\noindent
Kroupa, P. 2001, MNRAS, 322, 231

\noindent
Kroupa, P. 2003, New Astronomy Reviews, 8, 605

\noindent
Kroupa, P. 2004, New Astronomy Reviews, 48, 47.
% astro-ph 0309598.

\noindent
Kwan, J., \&  Scoville, N. Z. 1976, ApJ, 210, L39

\noindent
Lada, C. \& Lada, E. 2003, ARA\&A 41, 57

\noindent
Lai, S., Girart, J.~M., \& Crutcher, R.~M.\ 2003, \apj, 598, 392 

\noindent
Larson, R. B., \& Starrfield 1971, A\&A, 13, 190
%On the formation of massive stars and the upper limit of stellar masses. 

\noindent
Larson, R.B. 2003, Rep. Prog. Phys. 66, 1651 (astro-ph/0306595)

\noindent
Larson, R.B. 2005, in {\it IMF at 50: The Initial Mass Function
50 Years Later}, 

eds. E. Corbelli,   F. Palla, H. Zinnecker. Kluwer, in press

\noindent
Larwood, J. 1997, MNRAS, 290, 490
% "The tidal disruption of protoplanetary accretion discs"

\noindent
Lee, J.-K., \& Burton, M. G. 
2000, MNRAS, 315, 11

\noindent
Lombardi, J. C., Warren, J. S., Rasio, F. A., Sills, A., \& 
Warren, A. R. 
2002, ApJ, 568, 939

\noindent
Lombardi, J. C., Thrall, Ap. P., Deneva, J. S., Fleming, S. W.,
\& Grabowski, P. E. 

  2003, MNRAS 345, 762

\noindent
Lonsdale, C.~J., Becklin, E.~E., Lee, T.~J., \& Stewart, J.~M.\ 1982, AJ,
87, 1819

\noindent
Mac Low, M., \& Klessen, R. S. 
2004, Rev.Mod.Phys., 76, 125
% Control of star formation by supersonic turbulence

\noindent
Marti J., Rodriguez L.F., \& Reipurth B.
1998, ApJ, 502, 337

\noindent
McCaughrean, M. J. \& Mac Low, M. L. 1997, AJ, 113, 391

\noindent
McCaughrean, M. J. \& Stauffer, J. R. 1994, AJ, 108, 1382 

\noindent
McCaughrean, M. J., Stapelfeldt, K. R., \& Close, L. M. 2000,

 in: Protostars and Planets IV, p. 485

\noindent
McKee, C.~F.~\& Tan, J.~C.\
2002, Nature, 416, 59
% ``Massive star formation in 100,000 years from turbulent and
%   pressurized molecular clouds''

\noindent
Menten, K. M. \& Reid, M. J. 1995, ApJ, 445, L157

\noindent
Mermilliod, J.~\& Garc{\'{\i}}a, B.\ 2001, IAU Symposium, 200, 191 
% Spectroscopic Binaries in Young Open Clusters

\noindent
Meszaros, P. 2002, ARA\&A, 40, 137

\noindent
Moeckel, N. \& Bally, J. 2005, (in preparation).

\noindent
Morris, M. \& Serabyn, E. 1996, ARAA, 34, 645

\noindent
Nakano, T., Hasegawa, T., \& Norman, C. 1995, ApJ 450, 183

\noindent
Nielbock, M., Chini, R., J{\" u}tte, M., \& Manthey, E.\
2001, A\&A , 377, 273
% ``High mass Class I sources in M 17''

\noindent
O'Dell, C. R., Wen, Z., \& Hu, X. 1993, ApJ, 410, 696

\noindent
O'Dell, C. R. \& Wong, S. K. 1996, AJ, 111, 846

\noindent
O'Dell, C. R. 2001, ARA\&A, 39, 99O

\noindent
Owocki, S.P., Gayley, K.G., \& Cranmer, S.R. 1998, ASP Conf Ser 131,
Boulder-Munich II: 

Properties of Hot Luminous Stars, ed. I. Howarth (San
Francisco:  ASP), 237

\noindent
Padoan, P., \& Nordlund, A.  2002, ApJ, 576, 870

\noindent
Pfalzner, S. 2003, ApJ, 592, 986

\noindent
Plambeck, R. L., Wright, M. C. H., Mundy, L. G., \& Looney, L. W.
1995, ApJ, 455, L189

\noindent
Portegies Zwart, S. F., Makino, J., McMillan, S. L. W., \& Hut, P.
1999, A\&A, 348, 117

\noindent
Portegies Zwart, S. F., Makino, J., McMillan, S. L. W., \& Hut, P.
2002, ApJ, 565, 265

\noindent
Preibisch, T. et al. 1999, New Astronomy 4, 531

\noindent
Price, N. M. \& Podsiadlowski, Ph. 1995, MNRAS, 273, 1041 

\noindent
Proga, D., MacFadyen, A.~I., Armitage, 
P.~J., \& Begelman, M.~C.\ 2003, ApJ, 599, L5 
% Axisymmetric Magnetohydrodynamic Simulations of the 
% Collapsar Model for Gamma-Ray Bursts

\noindent
Reipurth, B., \& 
Bally, J.\ 2001, ARA\&A, 39, 403 

\noindent
Schertl, D., Balega, Y.~Y., Preibisch, T., \& Weigelt, G.\ 2003, \aap, 402, 267 
% Orbital motion of the massive multiple stars in the Orion Trapezium

\noindent
Schmeja, S., Klessen, R. S., \& Froebrich, D.
2005, A\&A (in press).

\noindent
Schultz, A.~S.~B., 
Colgan, S.~W.~J., Erickson, E.~F., Kaufman, M.~J., Hollenbach, D.~J., 

 O'Dell, C.~R., Young, E.~T., \& Chen, H.\ 1999, \apj, 511, 282 

\noindent
Scoville, N., Kleinmann, S.~G., Hall, D.~N.~B., \& Ridgway, S.~T.\ 1983, 
\apj, 275, 201 

\noindent
Shepherd, D. S., Yu, K. C., Bally, J., \& Testi, L.
2000, ApJ, 535, 833
% ``The Molecular Outflow and Possible Precessing Jet from the Massive Young
%			Stellar Object IRAS 20126+4104''

\noindent
Shu, F.~H., Najita, J.~R., 
Shang, H., \& Li, Z.-Y.\ 2000, Protostars and Planets IV, 

eds. V. Mannings, A. Boss, \& A. Russell p. 789 

\noindent
Smith, N., Davidson, K., Gull, T.R., Ishibashi, K., \& Hillier, D.J. 2003,
ApJ, 586, 432

\noindent
Soker, N., Regev, O., Livio, M., \& Shara, M. M.
1987, ApJ, 318, 760

\noindent
Stahler, S. W.,  Palla, F., \& Ho, P. T. P. 2000, Protostars and
Planets IV,

eds. V. Mannings, A. Boss, \& A. Russell p. 327

\noindent
Stolovy, S. R. et al. 1998, ApJ, 492, L151

\noindent
Stone, R. C. 1991, AJ, 102, 333

\noindent
Stone, J. M., Xu, J., \& Mundy, L. G. 1995, Nature, 377, 315

\noindent
Tan, J. C. 2004, ApJ, 607, L47
 %astro-ph/0401552)

\noindent
Tan, J.~C.~\& McKee, C.~F.\ 2002,
in {\it Hot Star Workshop III: The Earliest Phases of Massive 

  Starbirth} ed. P. Crowther, ASP Conf.Ser, 267, p. 267

\noindent
Toomre, A. \& Toomre, J. 1972, ApJ, 178, 623

\noindent
Torrelles, J.~M. et al. 2001a, Nature, 411, 277

\noindent
Torrelles, J.~M. et al. 2001b, ApJ, 560, 853

\noindent
van Altena, W. F., Lee, J. T., Lee, J. F., Lu, P. K., \& 
Upgren, A. R. 
1988, AJ, 95, 1744

\noindent
von Zeipel, H. 1924, MNRAS, 84, 665

\noindent
Watkins, S. J., Bhattal, A. S., Boffin, H. M. J., Francis, N., \&
Whitworth, A. P.

1998a, MNRAS, 300, 1205

\noindent
Watkins, S. J., Bhattal, A. S., Boffin, H. M. J., Francis, N., \&
Whitworth, A. P.

1998b, MNRAS, 300, 1214

\noindent
Weigelt, G., Balega, Y., Preibisch, T., Schertl, D.,
Sch{\" o}ller, M., \& Zinnecker, H. 

  1999, A\&A 347, L15

\noindent
Wiseman, J.~J.\& Ho, P. T. P. 1998, ApJ, 502, 676

\noindent
Wolfire, M. G. \& Cassinelli, J. P. 1986, \apj, 310, 207

\noindent
Wolfire, M. G. \& Cassinelli, J. P. 1987, \apj, 319, 850

\noindent
Woosley, S. E. 1993, ApJ, 405, 273
%Gamma-ray bursts from stellar mass accretion disks around black holes

\noindent
Yorke, H. W. \& Kr\"ugel, E. 1977, A\&A 54, 183

\noindent
Yorke, H. W., \& Sonnhalter, C. 2002,
ApJ, 569, 846

\noindent
Yorke, H. W. 2003, IAUS 221, 108 (Sydney)

\noindent
Yorke, H. W. 2002,
in {\it Hot Star Workshop III: The Earliest Phases of Massive Starbirth}

 ed. P. Crowther, ASP Conf.Ser, 267, p. 165

\noindent
Zinnecker, H.\ 1982, New 
York Academy Sciences Annals, 395, 226   

\noindent
Zinnecker, H., \& Bate, M. R. 2002,
in {\it Hot Star Workshop III: The Earliest Phases of Massive 

  Starbirth} ed. P. Crowther, ASP Conf.Ser, 267, p. 209

%  TABLES

%Table 1: Predictions of Massive Star Formation Mechanisms
\begin{deluxetable}{llll}
\tablewidth{0pt}
\tablecolumns{4}
\tablecaption{
Predictions of Massive Star Formation Mechanisms
 \label{table1}}
 \rotate
\tablehead{
  \colhead{Type of Study} &
  \colhead{Direct Accretion}  &
  \colhead{Mergers} &
  \colhead{Proposed Observation}
}
\startdata
Cores               &  isolated,  non-interacting
                    &  clustered, interacting
                    & cm, mm, sub-mm interferometry \\

Disks               &  stable, thin, accreting
                    &  transient, thick, expanding
                    &  IR, radio imaging \\

Outflows            &  collimated, quasi-steady
                    &  wide-angle eruptions
                    &  H$_2$, [FeII], CO, SiO, radio continuum \\

Massive YSOs        &  stable  IR,
                    &  flaring IR \& radio
                    &  thermal IR and radio continuum, \\

                    &  can be isolated
                    &  in dense clusters only
                    &  comparison with older data\\

Young Clusters      & low star density,
                    & high star density,
                    & wide-field IR imaging \\

                    & moderate stellar spin,
                    & fast stellar spin,
                    & \\

                    & small binary fraction
                    & high binary fraction
                    & \\

Remnant Associations &  low $\sigma _V$,
                    &  high $\sigma _V$,
                    & 2MASS, Spitzer, MSX archives \\

                    & few runaways till first SN
                    & runaways prior to first SN
                    & \\

\enddata

\raggedright
~\\
Notes:

\end{deluxetable}

\clearpage
% FIGURES

\begin{figure}
%\plotfiddle{Bally_mergers.fig1.eps}{4.0in}{0}{50}{50}{-170}{0}
\plotone{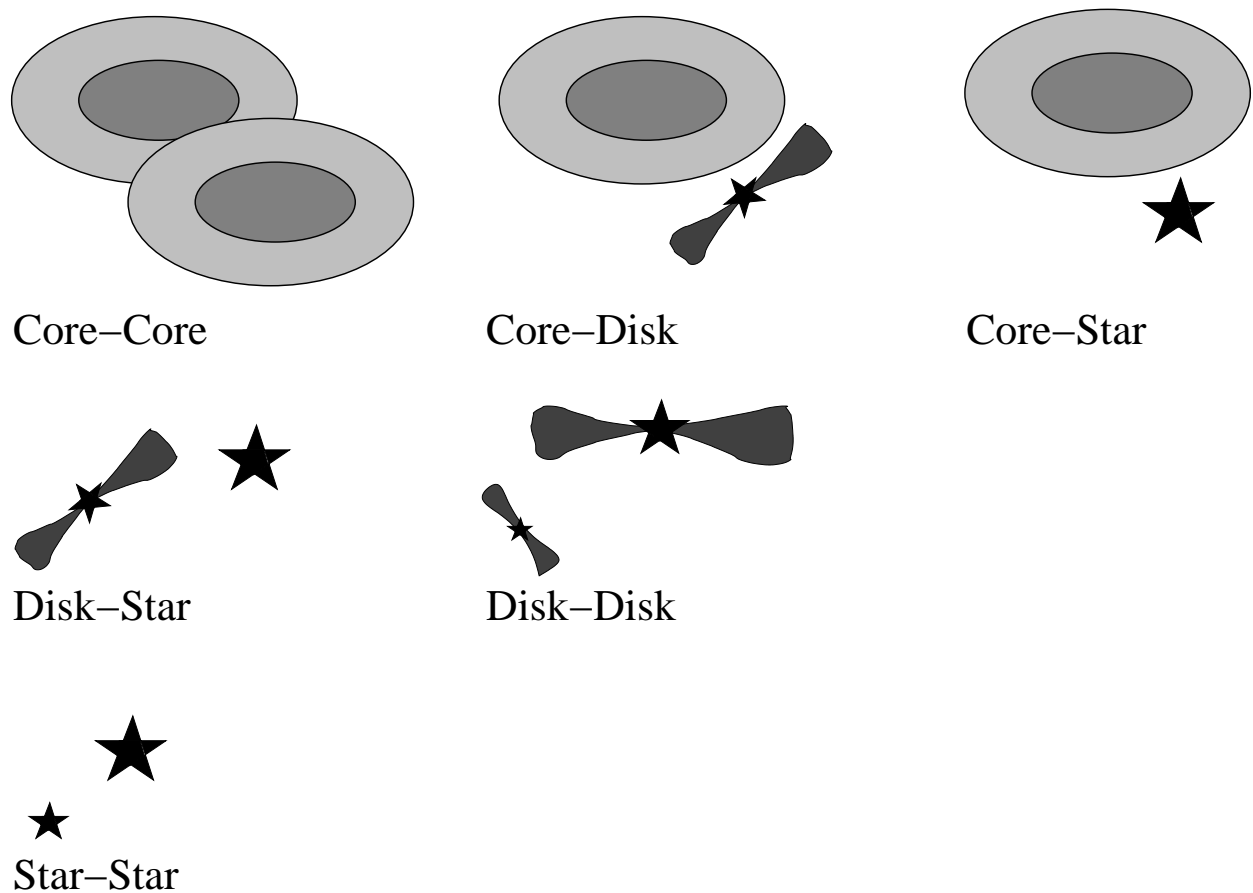}
\caption{
A cartoon showing six types of
interactions possible in a high density cluster environment.
It is assumed that there are three types of recognizable entity
in such an environment distinguished by increasing densities and 
diminishing dimensions; pre-stellar cloud cores, disks containing stars, 
and naked stars.  Each entity can interact with the other two.   
The interactions of cores, disks, or stars with starless cores tend to 
be low-energy, long-duration events.  This process is unlikely to
contribute to significant growth of massive stars due to the 
radiation effects.  Star-star interactions are 
likely to be extremely rare.  The most likely interactions which 
lead to merging are between disks or between disks and naked
stars.  As discussed in the text, these interactions may produce 
capture-formed binaries.   On-going accretion from a core, or 
subsequent interactions of the binary with other stars or disks 
are most likely to lead to merging.
}
\label{fig2}
\end{figure}

\end{document}